\documentclass{aastex63} %Astrophical Journal
\usepackage{amsmath}
\usepackage{amssymb}
\usepackage[normalem]{ulem}
\usepackage{appendix}
\usepackage{graphicx}

\begin{document}
%\shorttitle{GRO J2058+42 outburst in 2019 with AstroSat}
%\shortauthors{Mukerjee et al.}
%
\title{Strange Quark Stars in 4D Einstein-Gauss-Bonnet gravity}
\author{Ayan Banerjee}
\affiliation{Astrophysics and Cosmology Research Unit, University of KwaZulu Natal, Private Bag X54001, Durban 4000, South Africa}
\email{ayan\_7575@yahoo.co.in}
\author{Takol Tangphati}
\affiliation{Department of Physics, Faculty of Science, Chulalongkorn University, Bangkok 10330, Thailand}
\email{takoltang@gmail.com}
\author{Phongpichit Channuie}
\affiliation{College of Graduate Studies, Walailak University, Nakhon Si Thammarat, 80160, Thailand,}
\affiliation{School of Science, Walailak University, Nakhon Si Thammarat, 80160, Thailand,}
\affiliation{Research Group in Applied, Computational and Theoretical Science (ACTS), Walailak University, \\Nakhon Si Thammarat, 80160, Thailand}
\affiliation{Thailand Center of Excellence in Physics, Ministry of Higher Education, Science, Research and Innovation, \\Bangkok 10400, Thailand}
\email{channuie@gmail.com}

\begin{abstract}
The existence of strange matter in compact stars may pose striking sequels of the various physical phenomena. As an alternative to neutron stars, a new class of compact stars called strange stars should exist if the strange matter hypothesis is true. In the present article, we investigate the possible construction of the strange stars in quark matter phases based on the MIT bag model. We consider scenarios in which strange stars have no crusts. Then we apply two types of equations of state to quantify the mass-radius diagram for static strange star models performing the numerical calculation to the modified Tolman-Oppenheimer-Volkoff (TOV) equations in the context of $4D$ Einstein-Gauss-Bonnet gravity. It is worth noting that the GB term gives rise to a non-trivial contribution to the gravitational dynamics in the limit $D \to 4$. However, the claim that the resulting theory is of pure graviton was cast in doubt by several grounds. Thus, we begin our discussion with showing the regularized $4D$ EGB theory has an equivalent action as the novel $4D$ EGB in a spherically symmetric spacetime. We also study the effects of coupling constant $\alpha$ on the physical properties of the constructed strange stars including the compactness and criterion of adiabatic stability. Finally, we compare our results to those obtained from the standard GR.
\end{abstract}

\section{Introduction}\label{intro:Sec}

In modern gravity theories, higher derivative gravity (HDG) theories have attracted considerable attention,
as an alternative theories beyond GR. Among many impressive outcomes, HDG shows quite different aspects from that in four dimensions, and Einstein–Gauss–Bonnet (EGB) theory \citep{Lanczos:1938sf} is one of them. The EGB theory is a natural extension of GR to higher dimensions, which emerges as a low energy effective action of  heterotic string theory \citep{Zwiebach:1985uq} ( see the extended discussion in  Refs. \citep{Wiltshire:1985us,Boulware:1985wk,Wheeler:1986} for more information). As the string theory yields additional higher order curvature correction terms to the Einstein action  \citep{Callan:1985ia}. Interestingly, 
the EGB Lagrangian is a linear combination of Euler densities continued from lower dimensions, has been widely studied from astrophysics to cosmology.  The EGB theory which contains quadratic
 powers of the curvature is a special case of Lovelocks' theory of gravitation (LG) \citep{Lovelock,Lovelock:1972vz}
and is free of ghost. In $4 D$ spacetime EGB and GR are equivalent, as the Gauss-Bonnet (GB) term does not give any contribution to the dynamical equations.

According to the recent theoretical developments, Glavan and Lin \citep{Glavan:2019inb} proposed 
a 4-dimensional EGB gravity theory by rescaling the coupling constant $\alpha \to \alpha/(D -4)$, 
and then taking the limit $D \to 4$, a non-trivial black hole solution was found. In Ref. \citep{Guo:2020zmf}, authors have studied geodesic motions of timelike and null particles in the spacetime of
the spherically symmetric $4D$ EGB black hole. It was suggested that one
can bypass the Lovelock's theorem and the GB term gives rise to a non-trivial contribution to the gravitational
dynamics. However, it seems that regularization procedure was originally be traced back to 
Tomozawa \citep{Tomozawa:2011gp} with finite one-loop quantum corrections to Einstein gravity. 
 One can say that this interesting proposal has opened  up a new window 
for several novel predictions, though the validity of this theory is at present under debate and doubts. 
The spherically symmetric black hole solutions and their physical properties have been discussed \citep{Glavan:2019inb} that claims to differ from the standard vacuum-GR Schwarzschild BH. In the same framework static and spherically symmetric Gauss-Bonnet black hole was used to reveal many interesting features (for review, see, for instance, 
\cite{Ghosh:2020syx,Konoplya:2020juj,Kumar:2020uyz,li04,Kumar:2020xvu,Zhang:2020sjh,Liu:2020vkh,we03}). In \cite{Kumar:2020owy,NaveenaKumara:2020rmi} a rotating black hole solution has been found  using the Newman-Janis algorithm. They showed that the rotating black hole has an additional GB  parameter $\alpha$ than the Kerr black hole, and it produces deviation from Kerr geometry. However, it is well known \cite{Hansen:2013owa} that the Newmann-Janis trick is not generally applicable  in higher curvature theories. Thus, rotating solution still remains to be found in 4$D$ EGB gravity. Beside that geodesics motion and shadow \citep{Zeng:2020dco}, the strong/ weak gravitational lensing by black hole  \citep{Islam:2020xmy,Kumar:2020sag,Heydari-Fard:2020sib,Jin:2020emq}, spinning test particle \citep{zh03},  thermodynamics 
 AdS black hole \citep{sa03}, Hawking radiation \citep{Zhang:2020qam,Konoplya:2020cbv}, quasinormal modes \citep{Churilova:2020aca,Mishra:2020gce,ar04}, and wormhole solutions \citep{Jusufi:2020yus,liu20}, were extensively analyzed. It has attracted a great deal of recent attention, see \citep{Jusufi:2020qyw,Yang:2020jno,Ma:2020ufk,si20} for more. More recently, the study of the possible existence of thermal phase transition between AdS to dS asymptotic geometries in vacuum in the context of novel 4D Einstein-Gauss-Bonnet (EGB) gravity has been proposed in Ref.\citep{Samart:2020sxj}.

However, there are several works \citep{Ai:2020peo,Mahapatra:2020rds,Arrechea:2020evj,Gurses:2020ofy} debating  that the procedure of taking $D \to 4$  limit in \citep{Glavan:2019inb} may not be consistent. Let us mention a few examples. It was shown in Ref. \cite{Gurses:2020ofy} that there exists no four-dimensional equations of motion constructed from the metric alone that could serve as the equations of motion for such a theory. The four-dimensional theory must introduce additional degrees of freedom. In any case, it cannot be a pure metric theory of gravity. Moreover, from the perspective of scattering amplitudes, it was demonstrated that the limit leads to an additional scalar degree of freedom, confirming the previous analysis. Taking these issues together, the approach proposed by Glavan and Lin maybe incomplete. In other words, a description of the extra degree of freedom is required. Since then, several regularization schemes, e.g., see \cite{Lu:2020iav,Kobayashi:2020wqy,Fernandes:2020nbq}, have been proposed in order to overcome these shortcomings. In fact, Lu and Y. Pang \cite{Lu:2020iav} have shown that the Kaluza-Klein approach of the $D \to 4$ limit leads to a class of scalartensor theory that belongs to the Horndeski class. Thus, it is important to check the equivalence of the actions in the regularized and novel $4D$ EGB theory for the particular kind of static spherically symmetric spacetime. 

Nevertheless, the 4$D$ EGB gravity witnessed significant
attention that includes finding astrophysical solutions
and investigating their properties. In particular, the
mass-radius relations are obtained for realistic hadronic and for strange quark star EoS \citep{Doneva:2020ped}. Precisely speaking, we are interested to investigate the behaviour of compact star namely strange  quark  stars 
in regularized 4$D$ EGB gravity. Matter at densities exceeding that
of nuclear matter will have to be discussed in terms of quarks. As mentioned in Ref. \citep{Haensel1986} that for quark matter models massive neutron stars may exist in the form of strange quark stars. Usually the quark matter phase is modeled in the context of the MIT bag model as a Fermi gas of $u$, $d$, and $s$ quarks. At finite densities and zero or small temperature, quark matter can exhibit substantial rich phase structures resulting from different pairing mechanisms due to the coupling of color, flavor and spin degrees of freedom, see e.g. Refs. \citep{RAJAGOPAL2001,alford2007,Alford2002}. In addition, a variety of different condensates underlying fundamental descriptions may be plausible.

An expectation is that quark matter might play an important role in cosmology and in astrophysics, see e.g. \citep{Alford2004,far84}. On the one hand, in cosmology, it may provide an explanation of a source of density fluctuation and as a consequence of how galaxies form generated by the quark-hadron transition. On the other hand, in astrophysics, quark matter is an interplay between general relativistic effects and the equation of state of nuclear particle physics. These objects are present in the form of the stellar equilibrium including neutron stars with a quark core, super massive stars, white dwarfs and even strange quark stars. Nevertheless, in all possible applications of quark matter from cosmology and astrophysics, our lack of knowledge of the exact equation posses the main source of uncertainties in describing stars. In order to study the stable/unstable configurations and even other physical properties of stars, the realistic equations of state (EoS) have to be proposed. The color-flavor locked phase appearing in three flavor (up, down, strange) matter posses the importance of condensates \citep{alf01,raj01,lugo02,Steiner:2002gx} and is shown to be the asymptotic ground state of quark matter at low temperature \citep{Schafer:1999jg}. For instant, the authors of Ref.\citep{Banerjee:2020stc} studied a class of static and spherically symmetric compact objects made of strange matter in the color flavor locked (CFL) phase in 4$D$ EGB gravity. 

The structure of the present work is as follows: after the introduction in Sec.\ref{intro:Sec}, we quickly review how to derive the field equations in the context of 4$D$ EGB gravity and show that it makes a nontrivial contribution to  gravitational dynamics in 4$D$ in Sec.\ref{sec2}.  In Sec.\ref{sec3} we discuss a class of static and spherically symmetric compact objects invoking the equation of state parameters in quark matter phases invoking massless quark and cold star approximations. In Sec.\ref{sec4}, we discuss the numerical procedure used to solve the field equations. In the same section, we report the general properties of the spheres in terms of the massless quark and cold star approximations. We analyzed the energy conditions as well as other properties of the spheres, such as sound velocity and adiabatic stability. Finally, we conclude our findings in the last section.

%%%%%%%%%%%%%%%%%%%%%%%%%%%%%%%%%%%%%%%%%%%%%%
\section{A review of the regularized $4D$ EGB theory}
\label{sec20}
%%%%%%%%%%%%%%%%%%%%%%%%%%%%%%%%%%%%%%%%%%%%%%
In this section, we will take a short recap of the  regularization technique developed in Ref. \citep{Fernandes:2020nbq}, and applied it to the novel $4D$ EGB theory in order to find the regularized action. This regularization method leads to a well defined action which is free from divergences, and produces well behaved second-order field equations. The general action in the  framework of GB theory in $D$-dimensional spacetime, is given by
\begin{eqnarray}\label{action1}
 S= \frac{c^4}{16\pi G_D} \int_{\mathcal{M}}d^D x \sqrt{-g} \left(R + \hat\alpha \mathcal{L}_{\text{GB}} \right) + \mathcal{S}_{\text{matter}},
\end{eqnarray}
where $g$ denotes the determinant of the metric $g_{\mu\nu}$ and $\hat\alpha$ is the GB coupling constant. Since, $\mathcal{S}_{\text{matter}}$ is the action of the standard perfect fluid matter and the GB term is
\begin{eqnarray}\label{gravity field eq}
\mathcal{L}_{\text{GB}}= R^2-4R_{\mu\nu}R^{\mu\nu}+R_{\mu\nu\alpha\beta}R^{\mu\nu\alpha\beta}~.
\end{eqnarray}
Varying the action (\ref{action1}) results to the following equations of motion
\begin{equation}\label{GBeq}
G_{\mu\nu} + \hat{\alpha} H_{\mu\nu} = \frac{8 \pi G}{c^{4}} T_{\mu\nu}\,\,~~~~~{\rm where}\,\,T_{\mu\nu}= -\frac{2}{\sqrt{-g}}\frac{\delta\left(\sqrt{-g}\mathcal{S}_m\right)}{\delta g^{\mu\nu}},
\end{equation}
where $G_{\mu\nu}$ is the Einstein tensor and $H_{\mu\nu}$ is a tensor carrying the contributions from
the GB term, which yield 
\begin{eqnarray}
G_{\mu\nu} &=& R_{\mu\nu}-\frac{1}{2}R~ g_{\mu\nu},\nonumber\\
H_{\mu\nu} &=& 2\Bigr( R R_{\mu\nu}-2R_{\mu\sigma} {R}{^\sigma}_{\nu} -2 R_{\mu\sigma\nu\rho}{R}^{\sigma\rho} + R_{\mu\sigma\rho\delta}{R}^{\sigma\rho\delta}{_\nu}\Bigl)- \frac{1}{2}~g_{\mu\nu}~\mathcal{L}_{\text{GB}},\label{FieldEq}
\end{eqnarray}
where $R_{\mu\nu}$ is the Ricci tensor, $R_{\mu\sigma\nu\rho}$ is the Riemann tensor, and $R$ is the Ricci scalar, respectively. Note that for $D > 4$, the equation of motion (\ref{GBeq}) is the well-known Einstein-Gauss-Bonnet theory. However, in $D = 4$ the GB term  vanishes identically and hence the field equations (\ref{GBeq}) reduces to the Einstein's theory. However, if $\hat{\alpha}$ is rescaled as $\hat{\alpha} \to {\alpha \over D-4}$ and taking the limit $D \to 4$,  the GB gives nontrivial contributions to a well defined action principle in four dimensions. In what follows, we will show how this method works for
static and spherically symmetric spacetime and then investigate contributions of the GB term on a compact stellar object.

For the stellar configurations, we assume that the energy momentum tensor $T_{\mu\nu}$ is a perfect fluid matter source,
which is 
\begin{eqnarray}\label{emt}
T_{\mu\nu} = (\epsilon+P)u_{\nu} u_{\nu} + P g_{\nu \nu}, \label{em}
\end{eqnarray}
where $P=P(r)$ is the pressure, $\epsilon \equiv \epsilon(r)$ is the energy density of matter, and $u_{\nu}$ is a $D$-velocity. Here, we consider the $D$-dimensional spherically symmetric metric anstaz describing the interior of the star
\begin{eqnarray}\label{metric01}
    ds^2_{D} %&=& - e^{2\Phi(r)}c^{2}dt^2 + e^{2\Lambda(r)}dr^2 + r^{2}d\Omega_{D-2}^2 \nonumber\\
    &=& - W(r) dt^2 + H(r) dr^2 + r^{2}d\Omega_{D-2}^2, 
\end{eqnarray} 
where $d\Omega_{D-2}^2$ is the metric on the unit $(D-2)$-dimensional sphere and $W=W(r)$ and $H=H(r)$ are functions of redial coordinate $r$, respectively. In the limit $D \to 4$, considering the metric (\ref{metric01}) and Eq. (\ref{emt}) for the perfect fluid and obtain the components of $(t,t)$ and $(r,r)$ in the following
forms
\begin{eqnarray}
   \frac{8 \pi G}{c^{4}} \epsilon &=& \frac{1}{r^2} + \frac{1}{r H} \left( \frac{H'}{H} - \frac{1}{r} \right) 
   - \frac{\alpha (H - 1)}{r^4 H^3}\,\Big(H^2 -H - 2H' r\Big)\,, \label{DR1} \\ 
  \frac{8 \pi G}{c^{4}}  P &=& -\frac{1}{r^2} + \frac{1}{r H} \left( \frac{1}{r} + \frac{W'}{W} \right) 
  +\frac{\alpha (H - 1)}{r^4 H^2 W}\,\Big(W(H - 1) + 2 W' r\Big) \,,\label{DR2}
  %\\
   % \frac{dP}{dr} &=& - (\epsilon + P) \frac{d\Phi}{dr} .  \label{DRE3}
\end{eqnarray}
where primes denote derivative with respect to $r$. This new theory has stimulated a series of research
works concerning to cosmological as well as astrophysical solutions. 
Even though there are several criticisms against this
model, including the fact that the rescaling proposed substitutes
a vanishing factor with an undetermined one (see for
example \citep{Ai:2020peo,Gurses:2020ofy,Lu:2020iav,Kobayashi:2020wqy, Hennigar:2020lsl,Fernandes:2020nbq,shu}). 
In addition, several alternate \textit{regularizations} method have also been proposed including
the Kaluza–Klein-reduction  procedure \citep{Lu:2020iav,Kobayashi:2020wqy}, the conformal subtraction 
procedure \citep{Hennigar:2020lsl,Fernandes:2020nbq}, and ADM decomposition analysis \citep{Aoki:2020lig}. 
Thus, the regularization scheme is not unique. In what follows, we follow the approach as proposed in \citep{Fernandes:2020nbq,Lu:2020iav,Yang:2020jno} with the following form 
\begin{eqnarray}\label{action1}
 S &=&  \frac{c^4}{16\pi G}\int_{\mathcal{M}} \mathrm{d}^4 x \sqrt{-g} \Big[R+ \alpha \Big(4G^{\mu \nu}\nabla_\mu \phi \nabla_\nu \phi - \phi \mathcal{L}_{\text{GB}}+ 4\Box \phi (\nabla \phi)^2 + 2(\nabla \phi)^4 \Big] + \mathcal{S}_{\text{matter}},
\end{eqnarray}
which can be seen to be free of divergences and $\phi$ is a scalar function of the space-time coordinates.  Interestingly, the scalar acted as a Lagrange multiplier in the action allowing for the GB term itself to appear in the  $4D$ field equations. Hence, we clearly see that the GB term does not vanish in the limit of $D=4$, and hence it
has an effect on gravitational dynamics in $4D$. Moreover, the action (\ref{action1}) belongs to a subclass of the Horndeski gravity \citep{Horndeski:1974wa,Kobayashi:2019hrl}  with  $G_2=8 \alpha X^2-2\Lambda_0$, $G_3=8 \alpha X$, $G_4=1+4 \alpha X$ and $G_5 = 4 \alpha \ln X$ \big(where $X=-\frac{1}{2} \nabla_{\mu} \phi \nabla^{\mu} \phi$\big). Namely, the idea is to consider a Kaluza-Klein ansatz \citep{Lu:2020iav,Kobayashi:2020wqy}
\begin{eqnarray}
ds_D^2=ds_4^2+e^{2\phi}d\Omega_{D-4}^2,
\end{eqnarray}
or by conformal subtraction \citep{Hennigar:2020lsl,Fernandes:2020nbq}, where the subtraction background is defined under a conformal transformation ${\tilde g}_{ab} = e^{2\phi}g_{ab}$ and a counterterm, i.e., $-\alpha\int_{\mathcal{M}} \mathrm{d}^4\sqrt{-{\tilde g}}\,\tilde{\mathcal{G}}$, is added to the original action \citep{Fernandes:2020nbq}. In the above expression $d\Omega_{D-4}^2$ is the line element on the internal maximally symmetric space and $\phi$ an additional metric function that depends only on the external $D$
dimensional coordinates. Now, varying the action (\ref{action1}), one can obtain the gravitational field equations of the regularized $4D$
EGB theory is \citep{Fernandes:2020nbq}
\begin{eqnarray}\label{gravity field eq}
 G_{\mu \nu} =  \alpha \hat{\mathcal{H}}_{\mu \nu} + \frac{8\pi G}{c^4}T_{\mu \nu},
 \label{regularized-EFE}
\end{eqnarray}
where $\hat{\mathcal{H}}_{\mu\nu}$ is defined by 
\begin{eqnarray}\label{H}
\hat{\mathcal{H}}_{\mu\nu} &=& 2R\big(\nabla_\mu \nabla_\nu \phi - \nabla_\mu\phi \nabla_\nu \phi\big) + 2G_{\mu \nu}\Big(\big(\nabla \phi\big)^2-2\Box \phi\Big)+ 4G_{\nu \alpha} \big(\nabla^\alpha \nabla_\mu \phi -\nabla^\alpha \phi \nabla_\mu \phi\big)\nonumber\\
 &+& 4G_{\mu \alpha} \big(\nabla^\alpha \nabla_\nu \phi - \nabla^\alpha \phi \nabla_\nu \phi\big) + 4R_{\mu \alpha \nu \beta}\big(\nabla^\beta \nabla^\alpha \phi - \nabla^\alpha \phi \nabla^\beta\phi\big) + 4\nabla_\alpha\nabla_\nu \phi \big(\nabla^\alpha \phi \nabla_\mu \phi\nonumber\\
 &-& \nabla^\alpha \nabla_\mu \phi \big)+4 \nabla_\alpha \nabla_\mu \phi \nabla^\alpha\phi \nabla_\nu \phi - 4\nabla_\mu \phi \nabla_\nu \phi \Big(\big(\nabla \phi\big)^2+ \Box \phi \Big) +4\Box \phi\nabla_\nu \nabla_\mu \phi - g_{\mu \nu} \Big[ 2R\big(\Box \phi  - (\nabla \phi)^2\big)
\nonumber\\
&+& 4 G^{\alpha \beta} \big( \nabla_\beta \nabla_\alpha \phi -\nabla_\alpha \phi \nabla_\beta \phi \big)+ 2\big(\Box \phi \big)^2 
- \big( \nabla \phi\big)^4 + 2\nabla_\beta \nabla_\alpha\phi \big(2\nabla^\alpha \phi \nabla^\beta \phi - \nabla^\beta \nabla^\alpha \phi \big) \Big].
\end{eqnarray}
and by varying with respect to the scalar field, we get
\begin{eqnarray}\label{scalar field}
\frac{1}{8}\mathcal{L}_{\text{GB}} &=&  R^{\mu \nu} \nabla_{\mu} \phi \nabla_{\nu} \phi - G^{\mu \nu}\nabla_\mu \nabla_\nu \phi - \Box \phi (\nabla \phi)^2 +(\nabla_\mu \nabla_\nu \phi)^2- (\Box \phi )^2 - 2\nabla_\mu \phi \nabla_\nu \phi \nabla^\mu \nabla^\nu \phi\,.
\end{eqnarray}
The trace of the field equations (\ref{gravity field eq}) is found to satisfy
\begin{eqnarray}\label{GB3}
R+\frac{\alpha}{2} \mathcal{L}_{\text{GB}}= -\frac{8\pi G}{c^4}T.
\end{eqnarray}

The authors of \citep{Fernandes:2020nbq} argued that the trace of the field
equation is exactly same form as of the original $4D$ EGB theory.  In continuation with the  above it is also argued that there maybe a hidden scalar degree of freedom in the original theory. Note that when $\mathcal{L}_{\text{GB}}=0$, 
the scalar field equation (\ref{scalar field}) can be seen to be exactly equivalent, which means  that the counter term added to the action must vanish on-shell. In other words, an on-shell action that 
is identical to the action of the original theory, and thus classical evolution of the gravity-matter system is independent of the hidden scalar field \citep{Fernandes:2020nbq}. This claim needs to be scrutinized carefully as we do here. 

With the spherical coordinates (\ref{metric01}), the nonvanishing components of the gravitational field equations \citep{Fernandes:2020nbq} written in terms metric components, which are 
\begin{eqnarray}
   \frac{8 \pi G}{c^{4}} \epsilon &=& \frac{1}{r^2} + \frac{1}{r H} \left( \frac{H'}{H} - \frac{1}{r} \right) \nonumber\\
   &+& \alpha  \Bigg(\frac{6 H' \phi'}{r^2 H^3}-\frac{2 H'\phi' }{r^2 H^2}+\frac{H' W' \phi'^2}{H^3 W }+\frac{2 H' \phi'^3}{H^3}+\frac{6 H' \phi'^2}{r H^3}+\frac{4 \phi''}{r^2 H}-\frac{4 \phi''}{r^2 H^2}\nonumber\\
   &&\qquad -\,\frac{2 \phi'^2}{r^2 H} -\frac{2 \phi'^2}{r^2 H^2}-\frac{4 W' \phi'^2}{r H^2 W }-\frac{W'^2 \phi'^2}{H^2 W^2} -\frac{2 W'\phi'\phi''}{H^2 W}+\frac{\phi'^4}{H^2}-\frac{4 \phi'^2 \phi''}{H^2}-\frac{8 \phi' \phi''}{r H^2}\Bigg)\,, \label{tt-phi} \\ 
  \frac{8 \pi G}{c^{4}}  P &=& -\frac{1}{r^2} + \frac{1}{r H} \left( \frac{1}{r} + \frac{W'}{W} \right) \nonumber\\
  &+&\alpha  \Bigg(\frac{2 H'\phi'^3}{H^3}-\frac{H' W' \phi'^2}{H^3 W}+\frac{H'^2 \phi '^2}{H^4}-\frac{4 H' \phi'^2}{r H^3}-\frac{4 H'\phi' \phi''}{H^3}+\frac{6 W' \phi '}{r^2 H^2 W}-\frac{2 W' \phi '}{r^2 H W}\nonumber\\
  &&\qquad -\,\frac{2 \phi'^2}{r^2 H}+\frac{6 \phi '^2}{r^2 H^2}+\frac{6 W' \phi'^2}{r H^2 W}+\frac{2 W' \phi' \phi''}{H^2 W}+\frac{4 \phi''^2}{H^2}+\frac{3 \phi'^4}{H^2}-\frac{4 \phi'^2 \phi''}{H^2}+\frac{8 \phi' \phi''}{r H^2}\Bigg) \,.\label{rr-phi}
\end{eqnarray}
To proof an equivalent between the two theories, one must quantify the solution of $\phi$ making both of  theories to have the same solution. Thus it is convenient to eliminate the matter part by subtracting Eqs. (\ref{DR1}) and (\ref{tt-phi}) and Eqs. (\ref{DR2}) and (\ref{rr-phi}). The resultant field equations read
\begin{eqnarray}
0 &=& \frac{2 H'}{r^3 H^3}-\frac{2 H'}{r^3 H^2}+\varphi  \left(\frac{6 H'}{r^2 H^3}-\frac{2 H'}{r^2 H^2}\right)+\varphi ^2 \left(\frac{H' W'}{H^3 W}+\frac{6 H'}{r H^3}-\frac{2}{r^2 H}-\frac{2}{r^2 H^2}-\frac{W'^2}{H^2 W^2}-\frac{4 W'}{r H^2 W}\right)\nonumber\\
&+& \frac{2 \varphi ^3 H'}{H^3}-\frac{2}{r^4 H}+\frac{1}{r^4 H^2}+\varphi ' \left(\frac{4}{r^2 H}-\frac{4}{r^2 H^2}-\frac{4 \varphi ^2}{H^2} -\varphi  \left[\frac{2 W'}{H^2 W}+\frac{8}{r H^2}\right]\right)+\frac{\varphi ^4}{H^2}+\frac{1}{r^4}\,,\label{varphi-tt}
\\
0 &=& \varphi ^2 \left(\frac{H'^2}{H^4} -\frac{H' W'}{H^3 W}-\frac{4 H'}{r H^3}-\frac{2}{r^2 H}+\frac{6}{r^2 H^2}+\frac{6 W'}{r H^2 W}\right)+\frac{2 \varphi ^3 H'}{H^3}+\varphi ' \left(\varphi  \left[\frac{2 W'}{H^2 W}-\frac{4 H'}{H^3}+\frac{8}{r H^2}\right]-\frac{4 \varphi ^2}{H^2}\right)\nonumber\\
&+&\frac{2}{r^4 H}-\frac{1}{r^4 H^2}+\frac{2 W'}{r^3 H^2 W}-\frac{2 W'}{r^3 H W}+\varphi  \left(\frac{6 W'}{r^2 H^2 W}-\frac{2 W'}{r^2 H W}\right)+\frac{4 \varphi '^2}{H^2}+\frac{3 \varphi ^4}{H^2}-\frac{1}{r^4}\,,
\label{varphi-rr}
\end{eqnarray}
where the new variable, $\varphi$, is defined as $\varphi = \phi'$ and $\varphi' = \phi''$. To eliminate $\varphi'$ form the Eqs. (\ref{varphi-tt}) and (\ref{varphi-rr}), we rewrite the above expressions as
\begin{eqnarray}
0 &=& \left(\sqrt{H}- r \varphi - 1\right) \Bigg(r^3 H W'^2\varphi -r W' W \left[H' r^2 \varphi + 2 H^{3/2} r \varphi +2 H^2-2 H (3 r \varphi +1)\right]+  \left(\sqrt{H}+ r \varphi + 1\right) W^2\nonumber\\
&&\quad\quad\quad\quad\quad\quad\quad\quad\;\times\left[ H^2 - 8 \sqrt{H}^{3} +H \left(2 r \varphi-r^2 \varphi ^2 +7\right)+2 H' \sqrt{H} r-2 H' r (r \varphi +2) \right] \Bigg)\,,
\label{sol-varphi-tt}\\
0 &=& \frac{H-(r \varphi +1)^2}{\Big(W' r^2 \varphi +W \big[2 (r \varphi +1)^2-2 H\,\big]\Big)^2}\nonumber\\
&\times& \Bigg(W^3 \Big[H-(r \varphi +1)^2\Big] \Big[H^2 \left(4 H' r-9 r^4 \varphi ^4-12 r^3 \varphi ^3-10 r^2 \varphi ^2-4 r \varphi +3\right)
-4 H'^2 r^2 \nonumber\\
&&\quad\quad\; +\,4 H' H r \left(r^2 \varphi ^2-2 r \varphi -1\right)+3 H^4+H^3 \left(6 r^2 \varphi ^2+4 r \varphi -6\right)\Big]\nonumber\\
&&\quad\quad\; +\,2 r^3 H W'^2 W \varphi \Big[ + \left(3 r^3 \varphi ^3+6 r^2 \varphi ^2+11 r \varphi +4\right)-H' r^2 \varphi -H^2 (3 r \varphi +4) \Big]\nonumber\\
&&\quad\quad\; +\,2 r H W' W^2 \Big[H \Big(6 r^5 \varphi ^5+20 r^4 \varphi ^4+33 r^3 \varphi ^3 -2 r^2 \varphi  (H'-19 \varphi ) +19 r \varphi +4\Big)+2 H' r^2 \varphi +H^3 (3 r \varphi +4)\nonumber\\
&&\quad\quad\; -\,H^2 \left(9 r^3 \varphi ^3+22 r^2 \varphi ^2+22 r \varphi +8\right)\Big] +2r^5 H^2  W'^3 \varphi ^2\Bigg)\,.
\label{sol-varphi-rr}
\end{eqnarray}
To do so, one may see that the above field equations (\ref{sol-varphi-tt}) and (\ref{sol-varphi-rr})
are identities when $\big(\sqrt{H}- r \varphi - 1\big)$ equals to zero. Obviously, the above scalar field 
equation is an identity when the scalar field $\varphi$ satisfies the relation:
\begin{eqnarray}
\varphi = \frac{1}{r}\Big( \sqrt{H} -1 \Big)\,.\label{varphi-sol}
\end{eqnarray}
Thus, the moral of the story is that the two $4D$ EGB theories are equivalent 
 by substituting the solution of $\varphi(r)$ in (\ref{varphi-sol}) to the field equation in (\ref{regularized-EFE})
in the static spherically symmetric spacetime (\ref{metric01}).

%%%%%%%%%%%%%%%%%%%%%%%%%%%%%%%%%%%%%%%%%%%%%%%%%%%%%%%

%%%%%%%%%%%%%%%%%%%%%%%%%%%%%%%%%%%%%%%%%%%%%%%%%%%%%%
 \section{Basic equations of EGB gravity} \label{sec2}
%%%%%%%%%%%%%%%%%%%%%%%%%%%%%%%%%%%%%%%%%%%%%%%%%%%%%

In the previous section we recap the regularized $4D$ EGB gravity, and show that  the regularized $4D$ EGB theory is equivalent to the original one in a spherically symmetric spacetime. Thus we anticipate the use of novel $4D$ EGB gravity not to constitute an impasse in this work.  Here, we start by assuming the general action of EGB gravity in $D$-dimensions 
and also deriving the equations of motion for the underlying theory. For the moment we take the action as
\begin{equation}\label{action}
	\mathcal{I}_{G}=\frac{c^4}{16 \pi G}\int d^{D}x\sqrt{-g}\left[ R +\alpha \mathcal{L}_{\text{GB}} \right]
+\mathcal{S}_{\text{matter}},
\end{equation}
where all the notations and symbols have their usual meaning with the EGB Lagrangian is denined in $\mathcal{L}_{\text{GB}}$ Eq. (\ref{gravity field eq}). The corresponding field equations can be derived by varying action with respect to the metric tensor $g_{\mu\nu}$,
which is exactly same as (\ref{GBeq}). It is interesting to note that the trace of the Eq. (\ref{GBeq}) is
\begin{eqnarray}
g^{\mu\nu}\,\Big(G_{\mu\nu}+\alpha H_{\mu\nu}\Big) &=& -R + \alpha\big(2D-8\big) R_{\mu\nu}R^{\mu\nu}+\alpha\big(2-D/2\big)R^{2}+\alpha \big(2-D/2\big)R_{\mu\nu\sigma\rho}R^{\mu\nu\sigma\rho}\nonumber\\&=& -R-\frac{\alpha (D-4)}{2}\mathcal{L}_{\rm GB}.\label{FieldEqT}
\end{eqnarray}
It has thus been argued that for $D = 4$, the GB term has no effect on
gravitational dynamics. However, rescaling the GB dimensional coupling constant $\alpha$ according to $\alpha\rightarrow\alpha/(D-4)$, the trace of the field equation (\ref{GBeq}) yields
\begin{eqnarray}\label{GB3}
R+\frac{\alpha}{2}\mathcal{L}_{\rm GB} = -\frac{8 \pi G}{c^{4}}T\,,
\end{eqnarray}
which is exactly the same form as the trace of the field equations obtained from regularize the $4D$ EGB theory. In this way, the GB term can yield a non-trivial contribution to the gravitational dynamics even in four dimensions. Thus, we can conclude that the two $4D$ EGB theories are equivalent in the static spherically symmetric spacetime. On the same way authors in \citep{Lin:2020kqe} have shown that  the equivalence of these two theories in a cylindrically symmetric spacetime. For complete description of the compact star,  we use the regularization process (see Refs. \citep{Glavan:2019inb,Cognola:2013fva}) in which the spherically symmetric solutions are also exactly same as those of other regularised theories \citep{Lu:2020iav,Hennigar:2020lsl,Casalino:2020kbt,Ma:2020ufk}.

%%%%%%%%%%%%%%%%%%%%%%%%%%%%%%%%%%%%%%%%%%%%%%%%%%%%%%%%%%%%%%%%%%

The line element of the static and spherically symmetric metric describing 
a stellar structure in $4D$ EGB theory has the following form
\begin{eqnarray}
ds^2 &=& - e^{2\Phi(r)}c^{2}dt^2 + e^{2\Lambda(r)}dr^2 + r^{2}d\Omega^2\,.
\label{line-exp}
\end{eqnarray} 
Since, the above line element is equivalent to the line element in (\ref{metric01}) 
for $e^{2\Phi(r)}c^{2} = W(r)$ and $e^{2\Lambda(r)}=H(r)$. Finally,  the TOV equations for this 
theory of gravity are nothing else that $(tt)$, $(rr)$ and hydrostatic continuity equations (\ref {GBeq}) yield
\begin{eqnarray}\label{DRE1}
&& \frac{2}{r} \frac{d\Lambda}{dr} = e^{2\Lambda} ~ \left[\frac{8\pi G}{c^4} \epsilon - \frac{1-e^{-2\Lambda}}{r^2}\left(1-  \frac{\alpha(1-e^{-2\Lambda})}{r^2}\right)\right]\left[1 +  \frac{2\alpha(1-e^{-2\Lambda})}{r^2}\right]^{-1}, \\ 
&& \frac{2}{r} \frac{d\Phi}{dr} = e^{2\Lambda} ~\left[\frac{8\pi G}{c^4} P + \frac{1-e^{-2\Lambda}}{r^2} \left(1- \frac{\alpha(1-e^{-2\Lambda})}{r^2} \right) \right] \left[1 +  \frac{2\alpha(1-e^{-2\Lambda})}{r^2}\right]^{-1},\label{DRE2} \\
&& \frac{dP}{dr} = - (\epsilon + P) \frac{d\Phi}{dr}.  \label{DRE3}
\end{eqnarray} 
As usual, the asymptotic flatness imposes $\Phi(\infty)=\Lambda(\infty)=0$ while  the regularity at the center requires $\Lambda(0)=~0$. 

It is advantageous to define  the gravitational mass within the sphere of radius $r$, such that $e^{-2\Lambda} =1-\frac{2G m(r)}{c^2 r}$.  Now, we are ready to write the Tolman-Oppenheimer-Volkoff (TOV) equations in a form we want to use. So, using (\ref{DRE2}-\ref{DRE3}), we obtain
the modified TOV as
\begin{equation}
{dP \over dr} = -{G\epsilon(r) m(r) \over c^{2}r^2}\frac{\left[1+{P(r) \over \epsilon(r)}\right]\left[1+{4\pi r^3 P(r) \over c^{2}m(r)}-{2G\alpha m(r) \over c^{2}r^3}\right]}{\left[1+{4G\alpha m(r) \over c^{2}r^3}\right]\left[1-{2Gm(r) \over c^{2}r}\right]}. \label{e2.11}
\end{equation}
If we take the $\alpha \to 0$ limit, the above equation reduces to the standard TOV equation of GR. Replacing the last equality in Eq. (\ref{DRE1}), we obtain  the gravitational mass:
\begin{equation}
m'(r)=\frac{6  \alpha G m(r)^2+4 \pi  r^6 \epsilon (r)}{4 \alpha G r m(r)+c^2 r^4}, \label{e2.12}
\end{equation}
using the initial condition $m(0)=0$. Then we use
the dimensionless variables $P(r)=\epsilon_{0}{\bar P}(r)$ and $\epsilon(r)=\epsilon_{0}{\bar \epsilon}(r)$ and $m(r)=M_{\odot}{\bar M}(r)$, with $\epsilon_{0}=1\,{\rm MeV}/{\rm fm}^{3}$. As a result, the above two equations become 
\begin{eqnarray}
{d{\bar P}(r) \over dr} &=& -{G{\bar \epsilon}(r) M_{\odot}{\bar M}(r) \over c^{2}r^2}\frac{\left[1+{{\bar P}(r) \over {\bar \epsilon}(r)}\right]\left[1 + {4\pi r^3 \epsilon_{0}{\bar P}(r) \over c^{2}M_{\odot}{\bar M}(r)}-{2G\alpha M_{\odot}{\bar M}(r) \over c^{2}r^3}\right]}{\left[1+{4G\alpha M_{\odot}{\bar M}(r) \over c^{2}r^3}\right]\left[1-{2GM_{\odot}{\bar M}(r) \over c^{2}r}\right]}  \nonumber \\
&=& - \frac{c_1 {\bar \epsilon}(r) {\bar M}(r)}{r^2} \frac{\left[1+{{\bar P}(r) \over {\bar \epsilon}(r)}\right] \left[1+{c_2 r^3 {\bar P}(r) \over {\bar M}(r)}-{2 c_1 \alpha {\bar M}(r) \over r^3}\right] }{\left[1+{4 c_1 \alpha {\bar M}(r) \over r^3}\right]\left[1-{2 c_1 {\bar M}(r) \over r}\right]}, \label{e2.11d}
\end{eqnarray}
and
\begin{eqnarray}
M_{\odot}\frac{d{\bar M}(r)}{dr} &=& \frac{6  \alpha G M^{2}_{\odot}{\bar M}(r)^2 + 4 \pi  r^6 \epsilon_{0}{\bar \epsilon}(r)}{4 \alpha G r M_{\odot}{\bar M}(r) + c^2 r^4} \label{e2.12d} \nonumber \\
\frac{d{\bar M}(r)}{dr} &=& \frac{6 c_1 \alpha {\bar M}(r)^2 + c_2 r^6 {\bar \epsilon}(r)}{4 c_1 \alpha r {\bar M}(r) + r^4}, \label{mr}
\end{eqnarray}
where $c_1 \equiv \frac{G M_{\odot}}{c^2} = 1.474 \text{ km}$ and $c_2 \equiv \frac{4 \pi \epsilon_0}{M_{\odot} c^2} = 1.125 \times 10^{-5} \; \text{km}^{-3}$. The relationship between mass $M$ and radius $R$ can be straightforwardly illuminated using Eq. (\ref{mr}) with a given EoS. Therefore, the final two Eqs. (\ref{e2.11d}) and (\ref{mr}) can be numerically  solved for a given EoS $P=P(\epsilon)$. In the next section, we will discuss the  strange matter hypothesis.

\section{Equation of state and numerical techniques} \label{sec3}

To understand, what kind of matter compact stars may be built up from, assuming an EoS is the most important step, which encompasses all the information regarding the stellar inner structure. Here, we solve the hydrostatic equilibrium Eq. (\ref{e2.11d})-
(\ref{e2.12d}) numerically for a specific EoS, $\epsilon$ = $f(P)$, where $\epsilon$ is the energy density and $P$ is the pressure. Since each possible EoS, there is a unique family of stars, parametrized by, say, the central density and the central pressure. The standard procedure is to derive the expressions $P=f(\rho)$ and $\epsilon =g(\rho)$, with $\rho$ being the baryon density, and then obtain an $\epsilon-P$ pair for every value of $\rho$. Fitting a curve to this data results in the EoS.
\begin{figure}
    \centering
    \includegraphics[width = 7.5 cm]{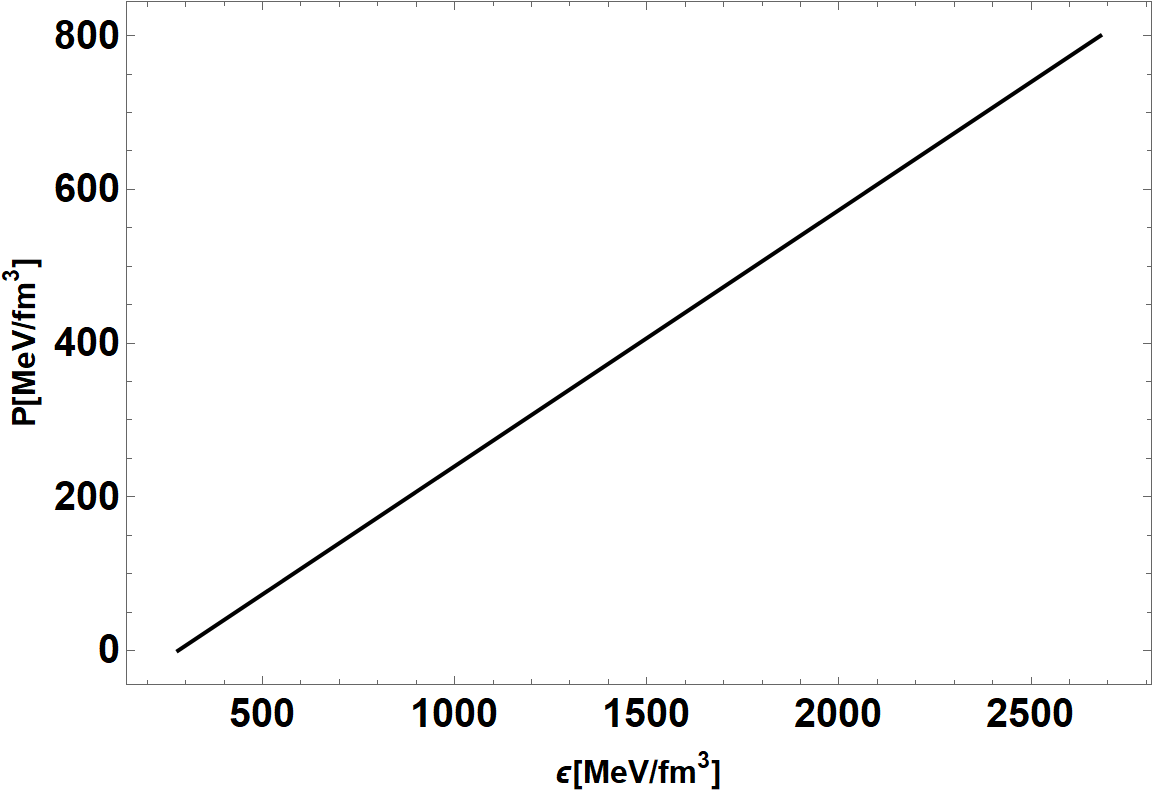}
    \includegraphics[width = 7.5 cm]{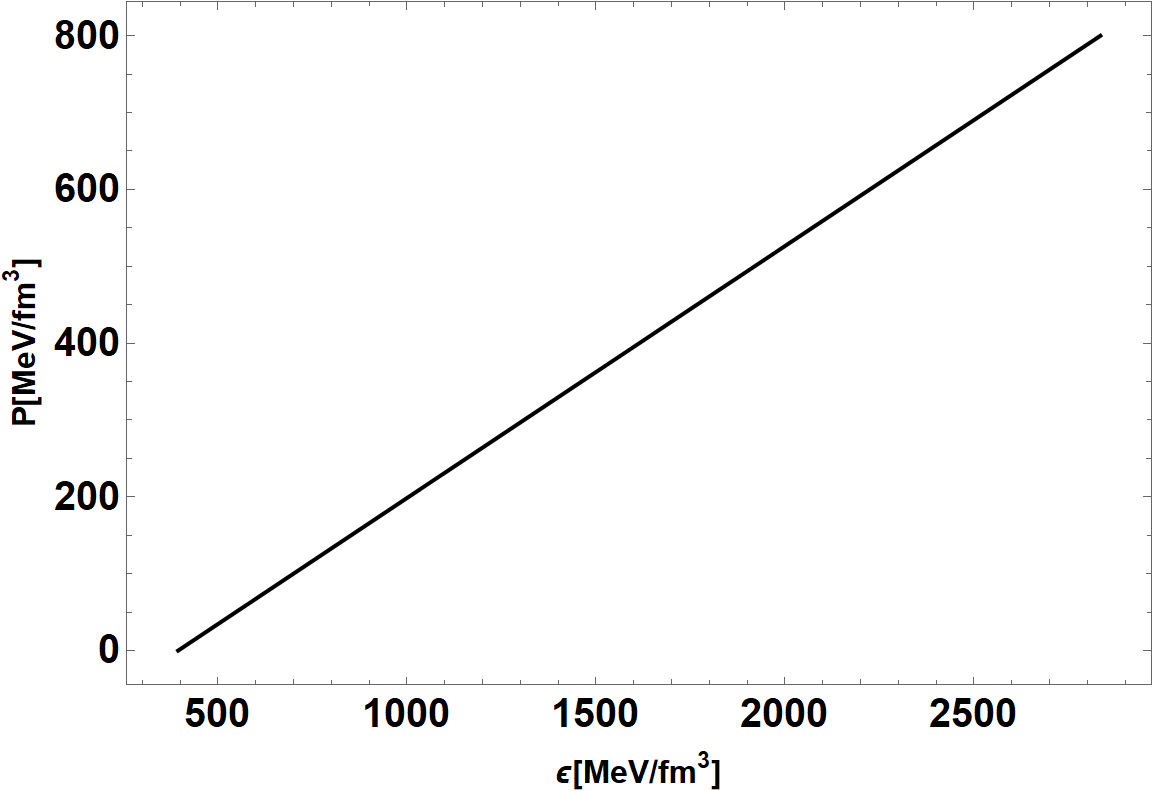}
    \caption{Figures demonstrate the relation between pressure and energy density for massless quark star (left panel) and cold star (right panel). For massless quark case, the EoS is given by $P = \frac{1}{3} \left( \epsilon - 4B \right)$ and the value of bag constant is about $B = 70 \text{ MeV} / \text{fm}^3$. For the cold star case, we use the mass of up quark, down quark, strange quark, electron and muon as 5, 7, 150, 0.5, and 105 MeV, respectively. With a consant $B = 70 ~\text{MeV}/\text{fm}^3$, the EoS reads $P = \frac{1}{3.05} \left( \epsilon - 368 \right)$.}
    \label{ff1}
\end{figure}
\subsection{Massless quark approximation}
Here, we begin with the discussion outlined above. One assumes that the asymptotically free quarks are confined in a finite region of space called a bag. The bag constant $B$ is basically considered as the inward pressure required to confine quarks inside the bag. It is usually given in a unit of energy per unit volume. In this particular model, we assume that the quark matter distribution is governed by the MIT bag EoS. For simplicity, it is assumed that $u$, $d$ and $s$ quarks are non-interacting and massless. Thus, according to the MIT bag model, the quark pressure $P$ is defined as
\begin{equation}\label{2.1}
P={\sum_{f=u,d,s}}{P^f}-{B},
 \end{equation}
where $P_f$ is the pressure due to each flavor. The energy density of each flavor $\epsilon_f$ is related to the corresponding pressure $P_f$ by the relation $P_{f}=\frac{1}{3}\epsilon_f$. The energy density due to the quark matter distribution in MIT bag model is governed by
\begin{equation}
{{\epsilon}}={\sum_{f=u,d,s}}{{\epsilon}_f}+B~. \label{2.2}
\end{equation}
By using Eqs.~\eqref{2.1} and \eqref{2.2} with the relation between $\epsilon_f$ and $P_f$, we end up with the well-known simplified MIT bag model and the EoS
takes the following simple form:
 \begin{equation}
\epsilon = 3P+4B\,.\label{EoS}
\end{equation}
 What we have to do next is to solve three equations with four unknown functions, which are $m(r),\,\Phi(r) ,\, P(r)\,$ and $\epsilon(r)$. Notice that the EoS for the massless quark approximations explicitly depends on the bag constant $B$ and the pressure $P(r)$. Due to the long range effects of confinement of quarks, the stability of strange quark star is essentially determined by the value of $B$, which can be seen from Fig.\ref{ff1} in the left panel. 
We then consider the customized TOV equations Eq. (\ref{e2.11d}) and mass function Eq. (\ref{mr}). Therefore, the mass is measured in the solar mass unit ($M_{\odot}$), radius in ${\rm km}$, while energy density and pressure are in ${\rm MeV}/{\rm fm}^{3}$.  In the present analysis, we treat the values of $B$ and $\alpha$ as a free constant parameters. Since, the parameter $B$ can vary from $57$ to $94 \,{\rm MeV}/{\rm fm}^{3}$ \citep{Witten}. For the study of quark matter with massless strange quarks,  we consider $B= 70 \,{\rm MeV}/{\rm fm}^{3}$.
\begin{figure}[!h]
    \centering
    \includegraphics[width = 7.5 cm]{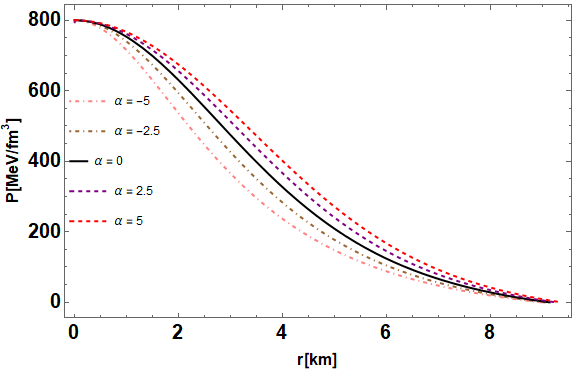}
    \includegraphics[width = 7.5 cm]{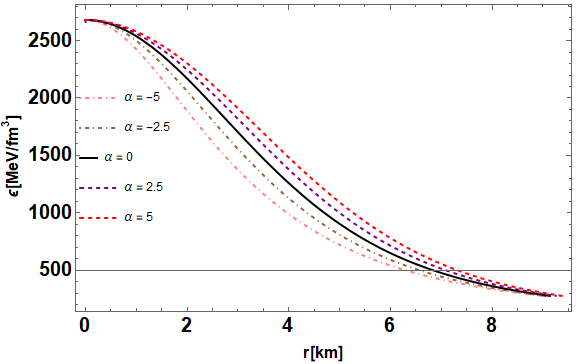}
    \caption{Variation of pressure (left panel) and the energy density (right panel) with radius for the strange quark stars 
 using the massless quark approximation for different values of $\alpha = 0, \pm 2.5~ \text{and} \, \pm 5\,{\rm  in~ km}^{2}\,$, where
we set $P(r_0) = 800.00 \, \text{MeV} / \text{fm}^3, B = 70.00 \, \text{MeV} / \text{fm}^3$, respectively.}
    \label{f1}
\end{figure}

\begin{figure}[!h]
    \centering
    \includegraphics[width = 7.5 cm]{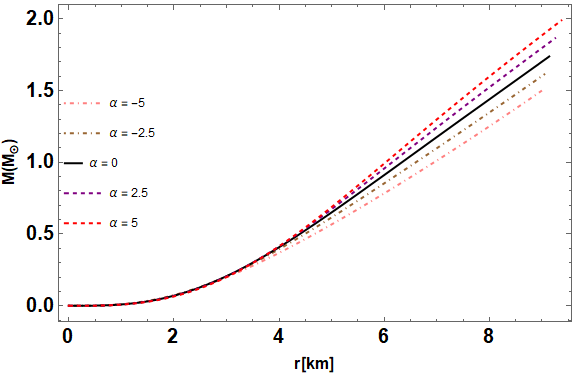}
    \includegraphics[width = 7.5 cm]{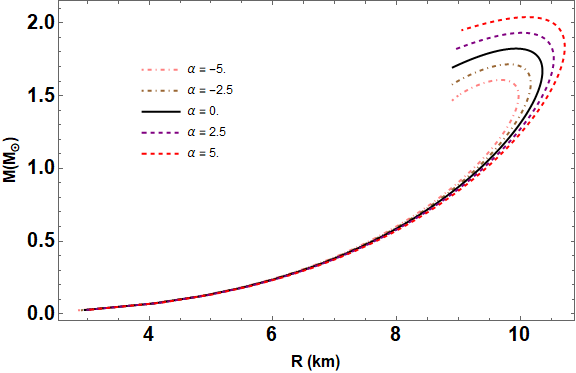}
    \caption{The mass-radius diagram using the massless quark approximation for  different values of $\alpha$; the notations are the same as in Fig. \ref{f1}.  The curve corresponds to  $\alpha = 0$ is representing GR case (solid black lines).}
    \label{f2}
\end{figure}

Given the set of differential equations (\ref{e2.11d}) and (\ref{mr}) together with the EoS (\ref{EoS}), we  apply numerical approach for integrating and calculate the maximum mass and other properties of the strange quark matter star. To do so, one can consider the boundary conditions $P(r_{0}) = P_{c}$ and $M(R) = M$, and integrates Eq. (\ref{e2.11d}) outwards to a radius $r = R$ in which fluid pressure $P$ vanishes for $P(R)=0$. This leads to the strange star radius $R$ and mass $M = m(R)$. The initial radius $r_{0}= 10^{-5}$ and mass $m(r_{0})= 10^{-30}$ are set to  very small numbers rather than zero to avoid discontinuities, as they appear in denominators within the equations.

We start from the center of the star for a certain value of central pressure,  $P({r_{0}})=800 \,{\rm MeV}/{\rm fm}^{3}$ and the radius of the star is identified when the pressure vanishes or drops to a very small value. For such a choice, we plot pressure and density versus distance from the center of strange star (see Fig. \ref{f1}). At that point we recorded the mass-radius relation of the star in Fig. \ref{f2}.  As one can see, the mass-radius ($M-R$) relation depends on the choice of the value of coupling constant $\alpha$. For $\alpha >0$ the mass of star for given radius increases with fixed value of $B$. In all the presented cases, one can note that there are significantly different for positive and negative values of  $\alpha$, but $\alpha =0$ case is equivalent
to pure general relativity. Moreover, as seen from Table \ref{ta11} and comparing the results to GR, one may obtain maximum mass for strange stars with positive $\alpha $. Therefore, we argue that a confirmed determination of a compact star with 2$M_{\odot}$,
which are actually very close to the ones of realistic neutron star models \citep{Haensel:1986qb}.

%%%%%%%%%%%%%%%%%%%%%%%%%%%%%%%%%%%%%
\subsection{Cold star approximation}
%%%%%%%%%%%%%%%%%%%%%%%%%%%%%%%%%%%%
This section contains a discussion of the zero temperature ($T=0$) and $m\neq0$.  Detailed calculations of the pressure, energy density, and baryon number density can be found, for example, in Ref. \citep{Glendenning2000}. In this scenario, we add the electrons to the system with their statistical weights ($=2$) due to the spin.  Performing the standard calculations, we obtain \citep{Glendenning2000}
\begin{eqnarray}
P &=& - B+\sum_{f} \Bigg[\frac{1}{4\pi^2}\Bigg(\mu_{f} k_{f}\Big(\mu_{f}^2 -\frac{5}{2}m_{f}^2  \Big) +\frac{3}{2} m_f^4 \ln\Big(\frac{\mu_{f}+ k_{f}}{m_f}\Big) \Bigg)\Bigg]\nonumber\\&&\quad\quad+\frac{1}{12\pi^2}\Bigg[\mu_{e} k_{e}\Big(\mu_{e}^2 -\frac{5}{2}m_{e}^2  \Big) +\frac{3}{2} m_e^4 \ln\Big(\frac{\mu_{e}+ k_{e}}{m_e}\Big) \Bigg], \label{eq37}\\
\epsilon &=& B+\sum_{f} \Bigg[\frac{3}{4\pi^2}\Bigg(\mu_{f} k_{f}\Big(\mu_{f}^2 -\frac{1}{2}m_{f}^2  \Big) +\frac{1}{2} m_f^4 \ln\Big(\frac{\mu_{f}+ k_{f}}{m_f}\Big) \Bigg)\Bigg]\nonumber\\&&\quad\quad+\frac{1}{4\pi^2}\Bigg[\mu_{e} k_{e}\Big(\mu_{e}^2 -\frac{1}{2}m_{e}^2  \Big) +\frac{1}{2} m_e^4 \ln\Big(\frac{\mu_{e}+ k_{e}}{m_e}\Big) \Bigg], \label{eq38}
\end{eqnarray}
where $k_{f}$ is the Fermi momentum for flavor $f$ with $k_f=\left(\mu_{f}^2-m_{f}^2 \right)^{1/2}$ and $k_e=\left(\mu_{e}^2-m_{e}^2 \right)^{1/2}$. Notice that there are four independent variables appeared in the above equations, i.e. $\mu_{u},\,\mu_{d},\,\mu_{s}$ and $\mu_{e}$. Strange stars are composed of $uds$ quarks. Hence, we constrain the chemical potentials of the quarks to a single independent variable $\mu$ such that $\mu_{d}=\mu_{s}=\mu$ and $\mu_{u}+\mu_{e}=\mu$. Thus, the two independent variables $\mu$ and $\mu_{e}$, two equations are necessary to produce a set of chemical potentials and solve the system for a pair of values for $\epsilon$ and $P$.

Knowing the quark chemical potentials, the relation between the pressure
and energy density of the quarks can be verified. The effects of the finite strange quarks mass on the energy density ($\epsilon$) and the pressure ($P$) for neutral quark matter including electrons showed that there was a sizable difference in the energy density and the pressure between zero strange quark mass and non-vanishing strange quark mass. However, the EoS in this case basically exhibits a non-linear behavior between $\epsilon$ and $P$, and as a result this non-linearity is very hard to be solved. This is because the quark chemical potentials increase when we increase the baryon number density, while the electron chemical potential
is negligible. We have thorough quantify the behaviors of the chemical potentials versus baryon number density using quark messes as given in Ref. \citep{Glendenning2000}.

Fortunately, it has been also noticed from Ref. \citep{Haensel}  that the resulting EoS $\epsilon=\epsilon(P)$ can be approximated by a non-ideal bag model which is written in the following form:
\begin{eqnarray}
    \epsilon = a\,B+b\,P, \label{app}
\end{eqnarray}
with $a$ and $b$ being arbitrary constants. In our case, we find that $\epsilon = 3.05\,P+368$, taking the value 
 $B=70\,{\rm MeV}/{\rm fm}^{3}$. Using the numerical calculations, the chemical potentials and the number density $\rho$ are simultaneously obtained. After substituting the results into Eq.~(\ref{eq37}) and Eq.~(\ref{eq38}), we finally end up with the EoS displaying a relationship between the energy density and pressure. However, the linear behavior is maintained by the relation given in Eq. \ref{app}, as illustrated in Fig.\ref{ff1} (right panel).

\begin{figure}[!h]
    \centering
    \includegraphics[width = 7.5 cm]{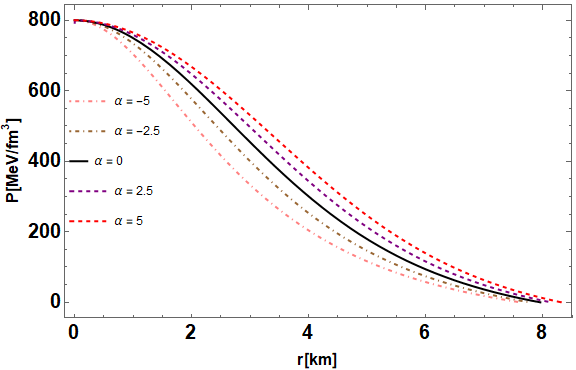}
    \includegraphics[width = 7.5 cm]{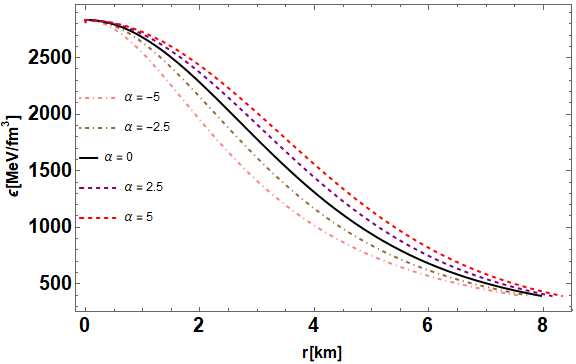}
    \caption{Variation of pressure (left panel) and the energy density (right panel) with radius for the strange quark stars using the cold star approximation for different values of $\alpha = 0, \pm 2.5~ \text{and} \, \pm 5\,{\rm  in~ km}^{2}\,$, where
we set $P(r_0) = 800.00 \, \text{MeV} / \text{fm}^3, B = 70.00 \, \text{MeV} / \text{fm}^3$, respectively.}
    \label{f3}
\end{figure}

\begin{figure}[!h]
    \centering
    \includegraphics[width = 7.5 cm]{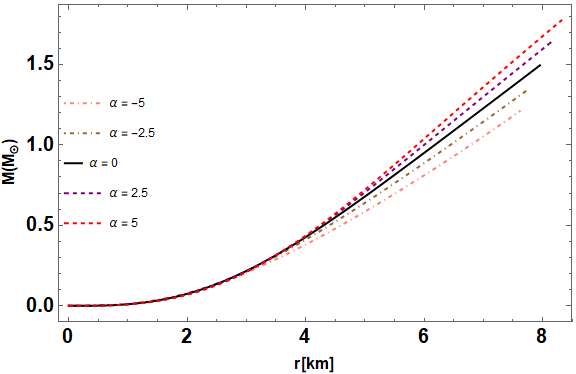}
    \includegraphics[width = 7.5 cm]{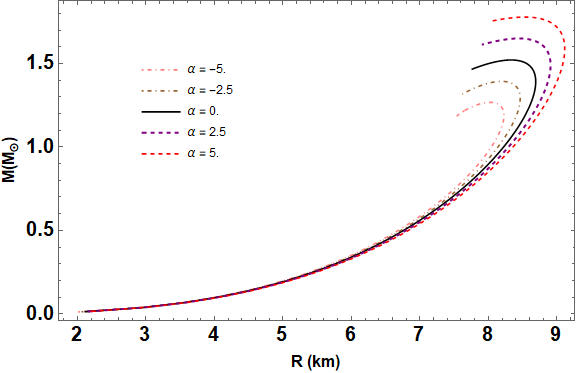}
    \caption{Unification diagram for the variation of mass as a function of radius (left panel) and the mass-radius curves for strange quark stars using the cold star approximation with different values of $\alpha$; the notations are the same as in Fig. \ref{f3}.}
    \label{f4}
\end{figure}

\begin{figure}
    \centering
    \includegraphics[width = 7.2 cm]{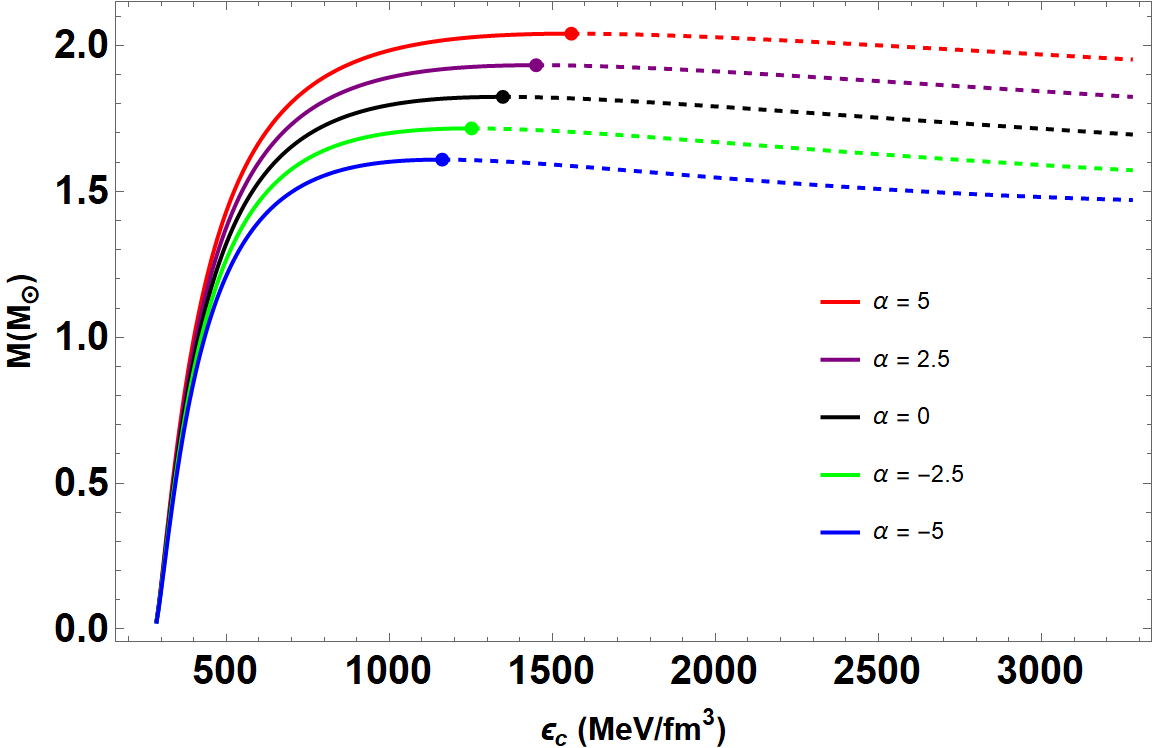}
    \includegraphics[width = 7.2 cm]{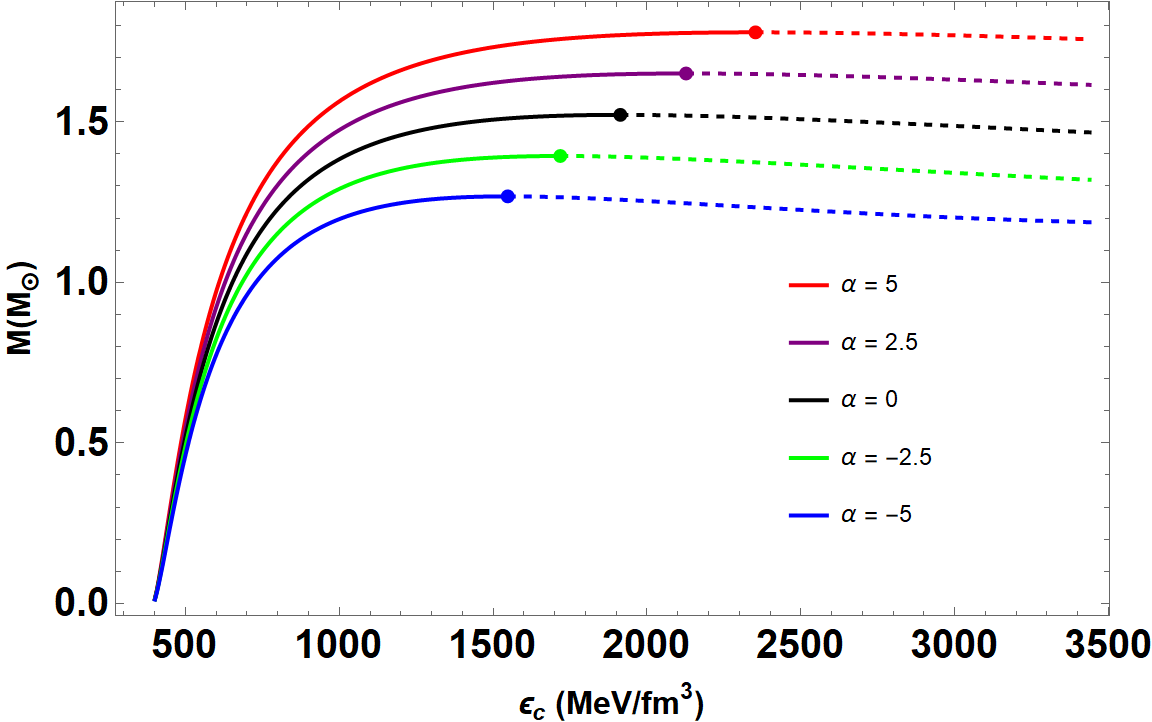}
    \caption{ Mass versus central density $\epsilon_{c}$ for compact star configurations obtained by solving the TOV equations (\ref{e2.11d}) and
(\ref{mr}) for all EoS introduced in Sect. \ref{sec3}.  The left panel is for massless quark star  while the right panel is for the cold star case. Here, solid lines represent positive slope of $dM/d\epsilon_{c}$, i.e., $dM/d\epsilon_{c}>0$. Stars on segments of the stellar sequence having positive slope are stable. The full circles represent the maximum mass configurations, $M_{\rm max}$.}
    \label{f5}
\end{figure}

\begin{figure}[!h]
    \centering
    \includegraphics[width = 12 cm]{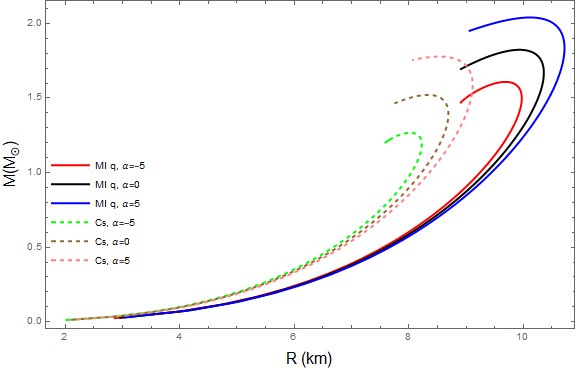}
    \caption{Figure displays the variation of mass (M) with star's radius (R) for the two types of EoSs: massless quark (Mlq) and cold stars (Cs).}
    \label{f6}
\end{figure}

The input data for the numerical calculation are similar to aforementioned.  The pressure and density versus radial distance from the center of cold star i.e.  quark matter at zero temperature are represented in Fig. \ref{f3}. All curves in Fig. \ref{f3}, note that the pressure and density are maximum at the center and decrease monotonically towards the boundary. In turn, to study the mass-radius relation and the mass vs. central density for cold quark matter EoS  are given for five representative values of $\alpha$ in Figs. \ref{f4} and \ref{f5}, respectively. For a given central density, the star mass grows with increasing  $\alpha$.  The maximum mass increases with increasing value of $\alpha$ and we find that, for  $\alpha=5\,$, the maximum mass becomes $M_{\mbox{max}}$ = 1.78 $M_{\odot}$. At that point we recorded the mass 
of the star, 1.52$M_{\odot}$ when $\alpha=0\,$ in GR.  For more clarity, the properties of stars with maximal mass are reported in Table \ref{ta11}  and are compared to GR ($\alpha=0\,$). Finally, in Fig. \ref{f6}, the mass-radius diagram is represented for two models (massless quark and  cold star approximation) for different values of the parameter $\alpha$. Recent discoveries of millisecond pulsar have shown that the neutron star mass distribution is much wider extending firmly up to $\sim 2 M_\odot$, and has already ruled out many soft EOSs. Fig. \ref{f6} clearly evidences that massless quark stars can achieve much higher masses and radii than cold star ones, which are actually very close to the realistic neutron star models with  $\sim 2 M_\odot$. While in Ref. \citep{Doneva:2020ped}, authors have obtained compact stars considering the hadronic and the strange quark star EoS. For the strange star EoS, the $M-R$ dependence is almost indistinguishable from GR for small masses and larger deviations exist only close to the maximum mass, which is similar to our solution. We discuss here the case depending on $\alpha =5$ and the value of $B=70 \, \text{MeV} / \text{fm}^3$, as an example, showing the results in Fig. \ref{f6} that if we set a limit on the maximum mass of a compact star, the corresponding maximum mass in the GR case cannot be achieved from the  same values. Interestingly, all values of $M_{\mbox{max}}$ for massless quarks  are higher than Chandrasekhar limit, which is about 1.4$M_{\odot}$.

\begin{figure}[h]
    \centering
    \includegraphics[width = 7.5 cm]{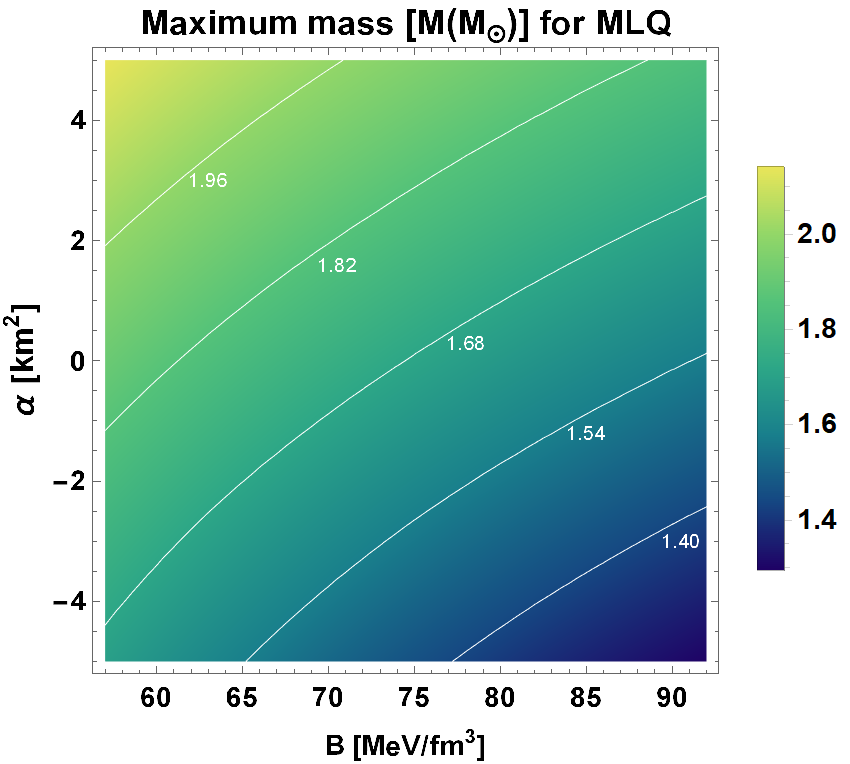}
    \includegraphics[width = 7.5 cm]{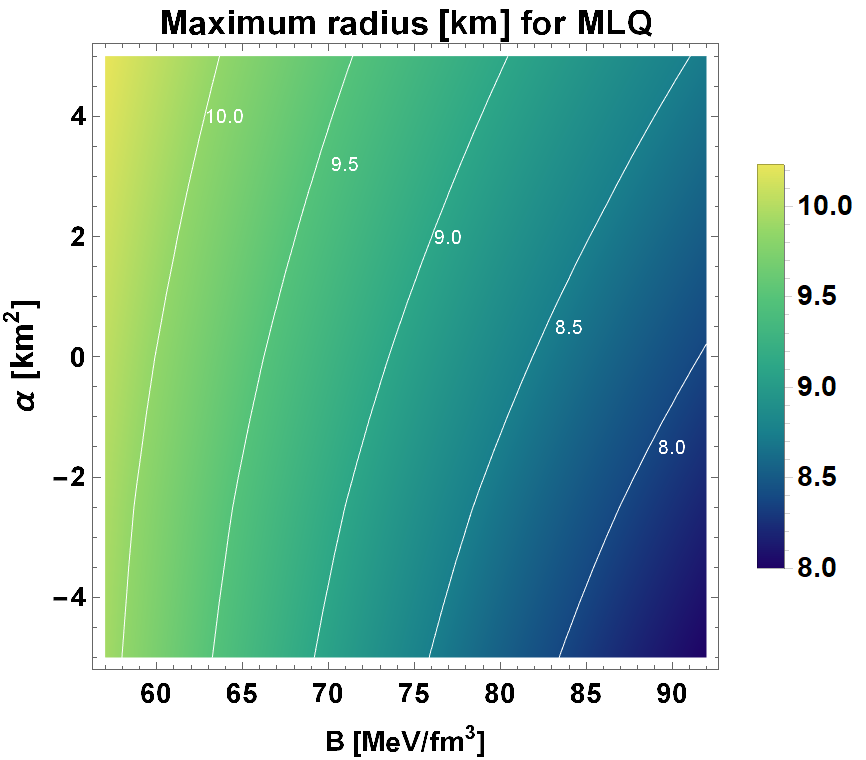}
    \caption{Figure shows the maximum masses (left panel) and their corresponding radii (right panel) for values of $P(r_0) = 700 \text{MeV/fm}^3$, $-5\leq \alpha \leq 5$ and $57\,{\rm MeV}/{\rm fm}^{3}\leq B\leq 92\,{\rm MeV}/{\rm fm}^{3}$ for the massless quark stars (MLQ). The white lines are equipped masses and radii lines.}
    \label{contour_MLQ}
\end{figure}

\begin{figure}[h]
    \centering
    \includegraphics[width = 7.5 cm]{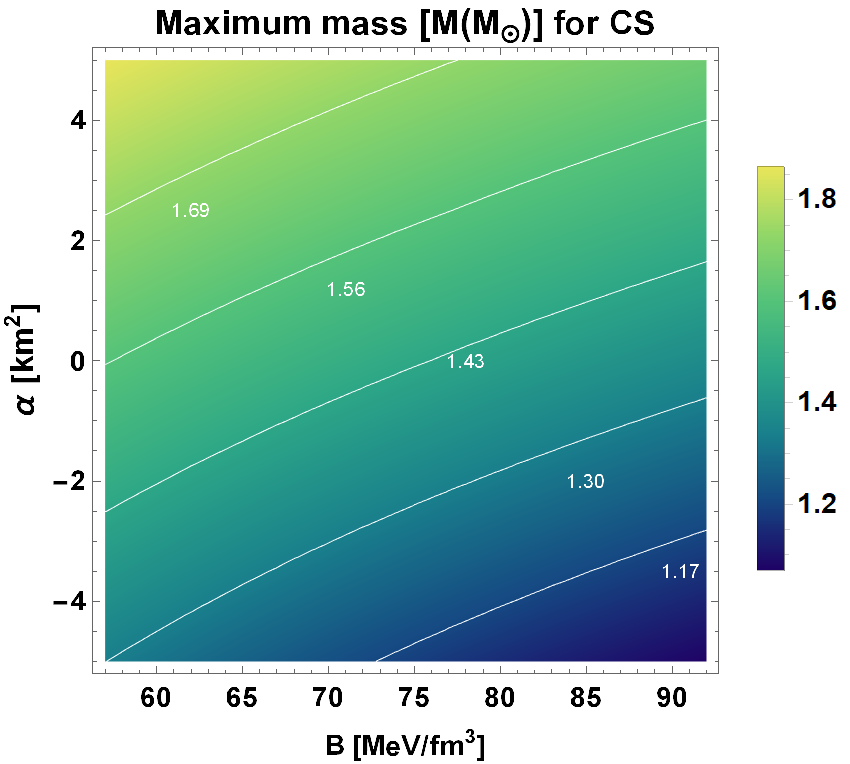}
    \includegraphics[width = 7.5 cm]{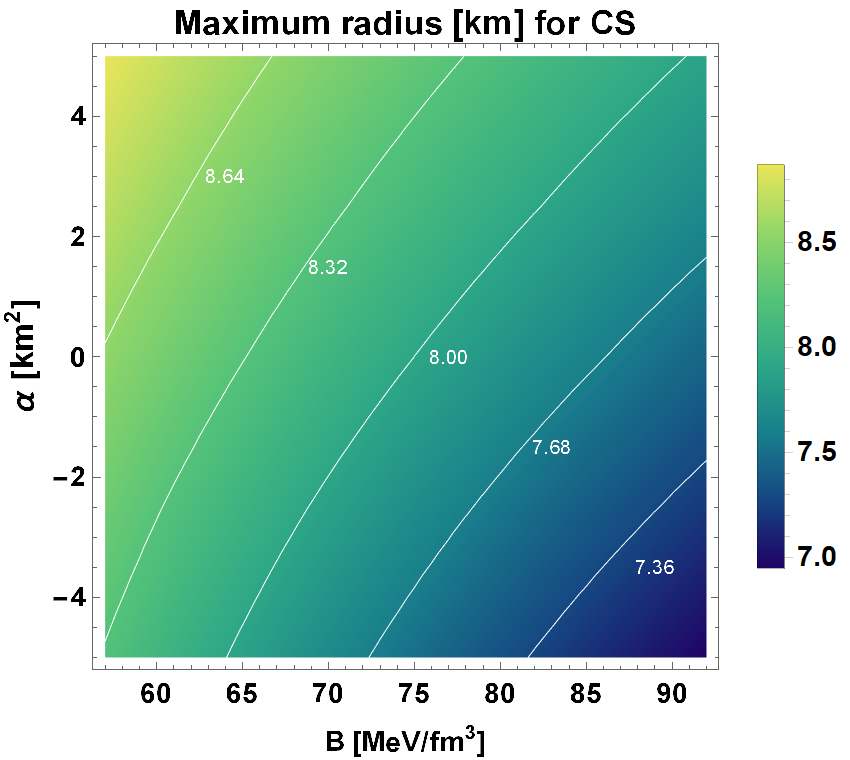}
    \caption{Figure shows the maximum masses (left panel) and their corresponding radii (right panel) for values of $P(r_0) = 700 ~\text{MeV/fm}^3$, $-5\leq \alpha \leq 5$ and $57\,{\rm MeV}/{\rm fm}^{3}\leq B\leq 92\,{\rm MeV}/{\rm fm}^{3}$ for the cold stars (CS). The white lines are equipped masses and radii lines.}
    \label{contour_CSS}
\end{figure}

In addition, a profile of solutions that covers the full range of values for $\alpha$ and the bag constant $B$ is presented in Figs  \ref{contour_MLQ} and \ref{contour_CSS}. With our parameter choice of $\alpha$ has a significant influence on the maximum masses and radius relation. In this case the maximum masses of QSs monotonically increasing with increasing $\alpha$ values. Observational constraints on GB constant were explored in Ref. \cite{Clifton:2020xhc}, which is $0 \lesssim \alpha \lesssim 10^2~ \text{km}^{2}$  based on observations of binary black holes. By analyzing Figs.\ref{contour_MLQ} for the massless quark star, it can be understood that, one may achieve the maximum mass above $M_{\rm max}\sim 2 M_{\odot}$ for $2.0<\alpha<3.0 \,{\rm km}^{2}$ and $B<60\,{\rm MeV/fm^{3}}$. Additionally, using Fig.\ref{contour_CSS} for the cold star, we can obtain the maximum mass above $M_{\rm max}\sim 2 M_{\odot}$ for $\alpha>5.0 \,{\rm km}^{2}$ and $B<60\,{\rm MeV/fm^{3}}$. The theory and findings suggest that our proposed model is in agreement with the current results for positive values of $\alpha$.

%%%%%%%%%%%%%%%
\begin{table}[ht!]
\begin{center}
\begin{tabular}{|c|c|c|c|c|c|c|}
\hline\hline
\multicolumn{4}{l}{\hskip 2cm \mbox{Massless Quark}}  & \multicolumn{3}{l}{\hskip 1.5cm \mbox{Cold Star} } \\
\hline\hline
$\alpha$ & $M_{\rm max}$ &  $R$  & $\epsilon_c$ & $M_{\rm max}$ &  $R$  & $\epsilon_c$ \\
 {\rm km}$^{2}$ &  $(M_{\odot})$ & (km) & (MeV/fm$^3$) &    $(M_{\odot})$ & (km) & (MeV/fm$^3$) \\
  \hline
  -5.0 & 1.61  & 9.68 & 1.16$\times 10^{3}$  & 1.27  & 8.02 & 1.55$\times 10^{3}$  \\
  -2.5 & 1.72 & 9.82 & 1.25$\times 10^{3}$  & 1.39 & 8.19 & 1.72$\times 10^{3}$ \\
     0 & 1.82 & 9.93 & 1.35$\times 10^{3}$ & 1.52 & 8.33 & 1.91$\times 10^{3}$ \\
  2.5 & 1.93 & 10.03 & 1.45$\times 10^{3}$  & 1.65 & 8.44 & 2.13$\times 10^{3}$ \\
  5.0 & 2.04 & 10.12 & 1.56$\times 10^{3}$ & 1.78 & 8.53 & 2.35$\times 10^{3}$ \\
  \hline\hline
\end{tabular}
\caption{We summarize the parameters of the strange quark stars using various values of the 4D EGB coupling constant, $\alpha$. Using Fig.\ref{f5}, we recorded the maximum mass of the stars $M$ in a unit of the solar mass $M_{\odot}$ with their radius $R$ in km and the central energy density $\epsilon_c$ for both the massless quark model and the cold star approximation.}\label{ta11}
\end{center}
\end{table}
 
%%%%%%%%%%%%%%%

\section{Structural properties of strange stars }\label{sec4}

For completeness, we would also like to show that the equations of stellar structure admit stable solutions and explore the physical properties in the interior of the  fluid sphere. In the following, we discuss the compactness and the stability of stars.

\subsection{Compactness}
\begin{figure}[h!]
    \centering
    \includegraphics[width = 7.5 cm]{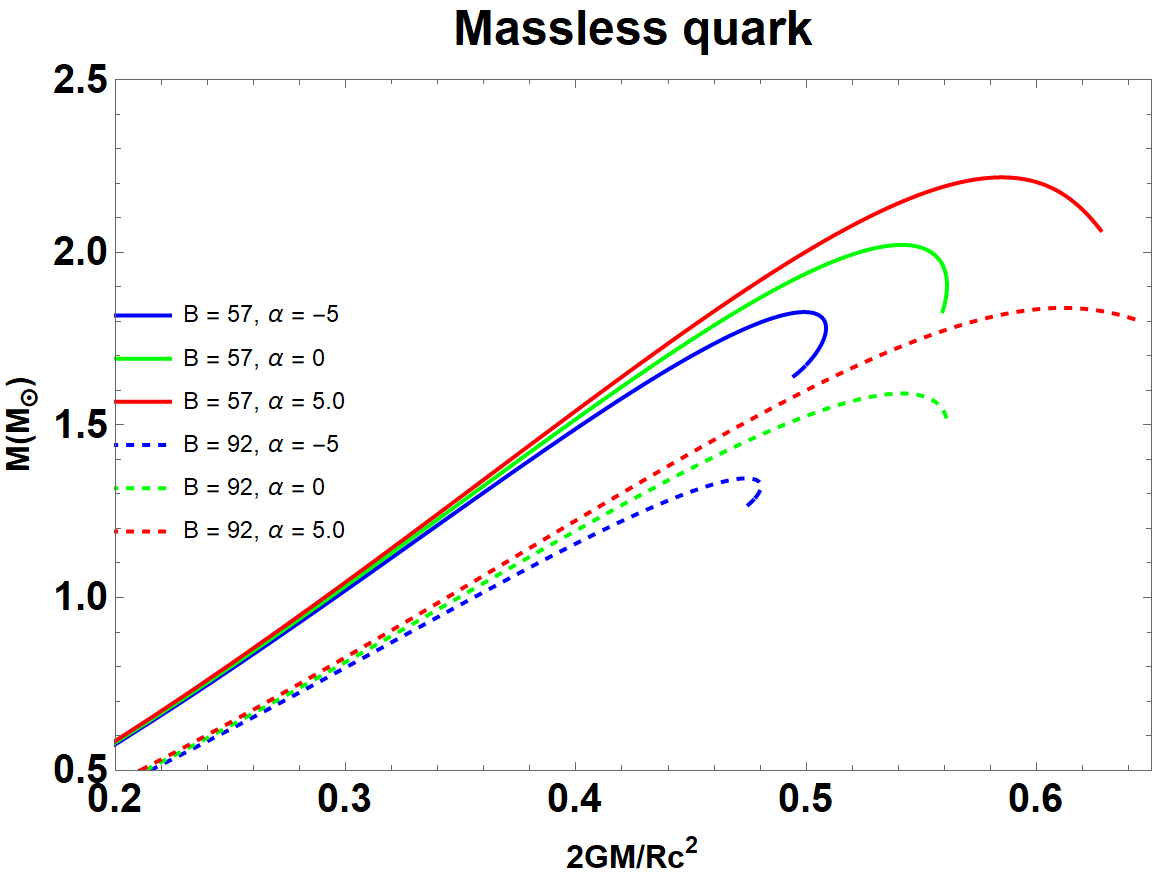}
    \includegraphics[width = 7.5 cm]{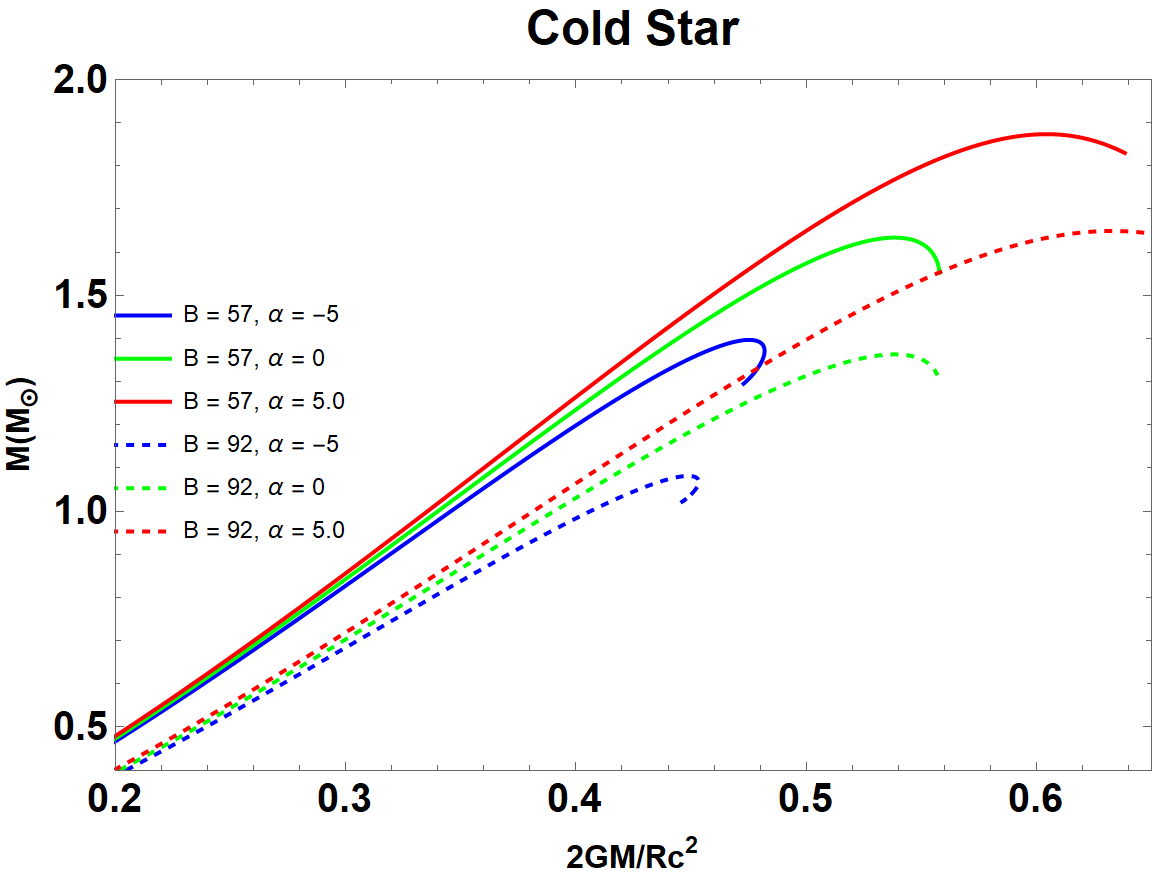}
    \caption{Figures represent the compactness of stars which is the relation of $r_{g}/R$. Here we display the mass $M$ and $r_{g}/R$. The left panel is the compactness of a massless quark with the Bag constants, $B = 57 \, \text{MeV} / \text{fm}^3$ and $B = 92 \, \text{MeV} / \text{fm}^3$, respectively. The right panel is the compactness of a cold star with the same set of parameters.}
    \label{f9}
\end{figure}

The qualitative effect of the compactness $2MG/Rc^{2}$ for each EoS is illustrated in Fig. \ref{f9}
for particular values of the bag constant $B$ and coupling constant $\alpha$.  Regarding the stars with the same $M$ and $\alpha$, it is noticed that the compactness decreases when the bag constant increases. Additionally, at the same values of $M,\,B$ and $\alpha$, the compactness in case of the cold star is higher than that of the massless quark EoS. Let us next define the Schwarzschild radius $r_{g}=2GM/c^{2}$. Interestingly, as mentioned in Ref. \citep{Haensel}, the compactness parameter $r_{g}/R$ characterizes the importance of relativistic effects for a star of mass $M$ and radius $R$. Figs. \ref{f9} show that the trend of stellar compactness lies in the range 
$0.5<r_{g}/R<0.6$ for both stars corresponding their respective EoS.

\begin{figure}[h!]
    \centering
    \includegraphics[width = 7.5 cm]{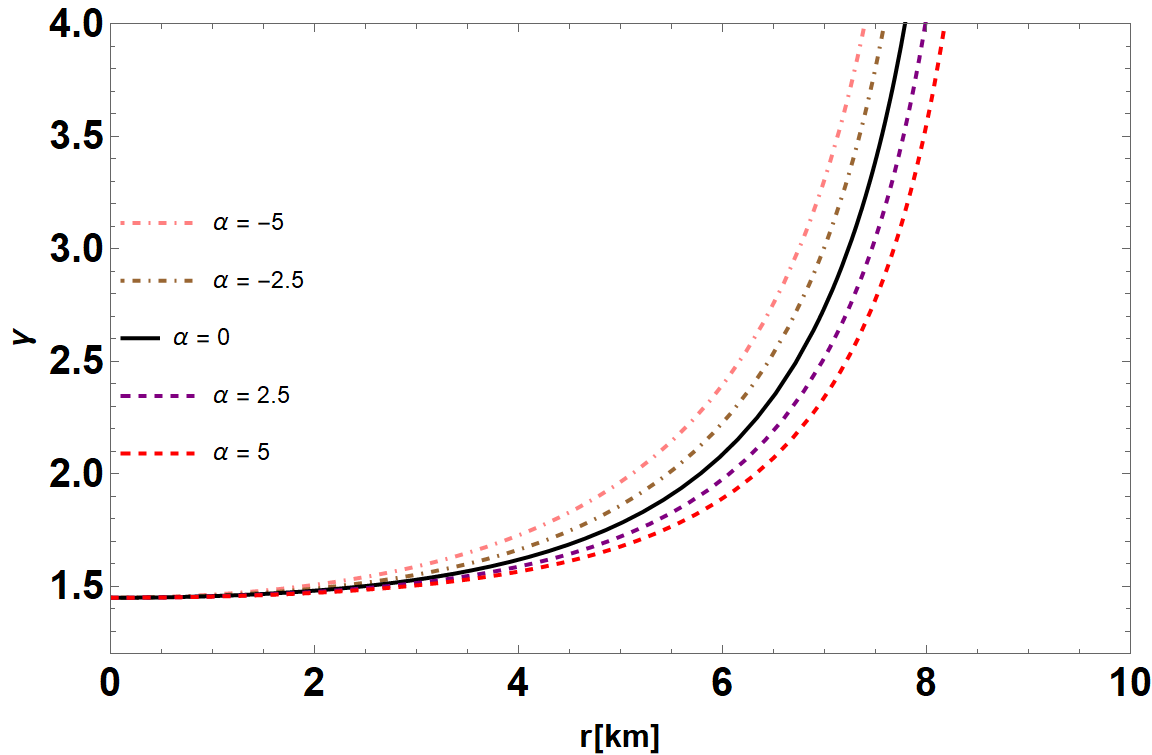}
    \includegraphics[width = 7.5 cm]{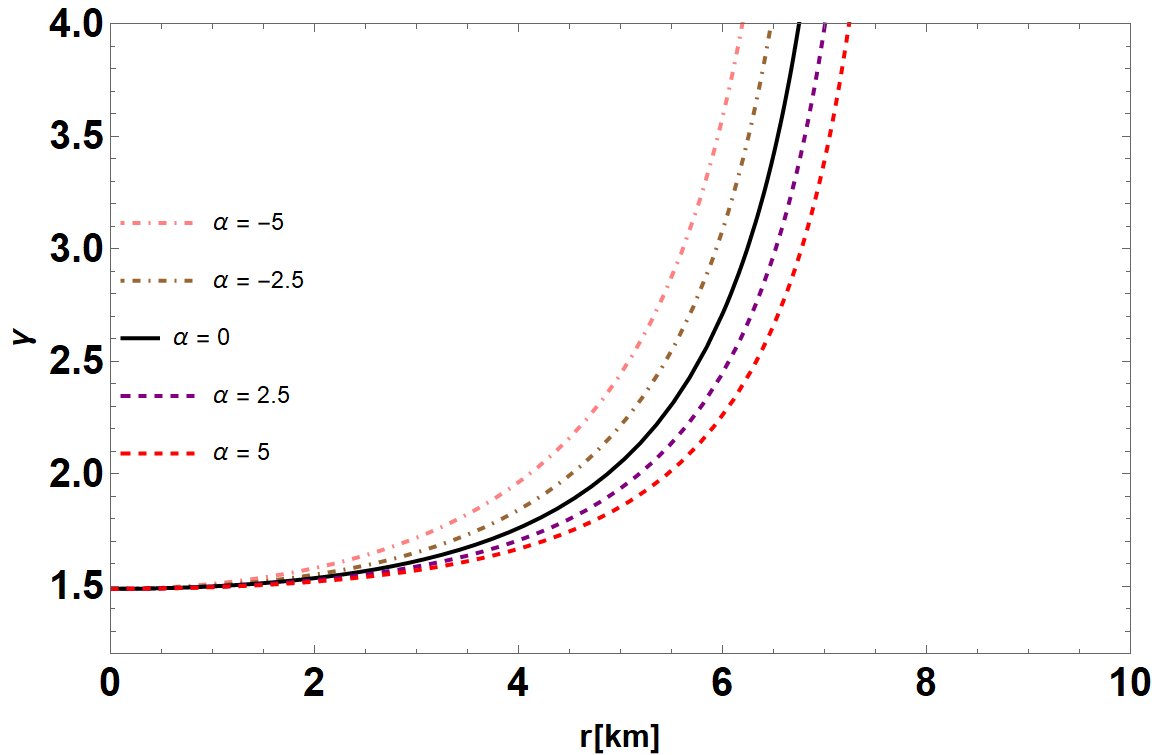}
    \caption{Plots for adiabatic index, $\gamma$, of the massless quark stars (left panel) and cold stars (right panel).}
    \label{gamma_MLQ_CSS}
\end{figure}

\subsection{Stability Test}

Our particular interest is to study the stability of the compact strange star.
A necessary (but not sufficient) condition for stability
of a compact star is that the total mass is an increasing
function of the central density, $dM/d\epsilon_c >0$ \citep{Arbanil:2016wud,Glendenning:1998ag}.
In Fig. \ref{f5}, we plot the dependence of masses of  compact stars in solar units on
their central density for massless quark (left panel) and cold star case (right panel).
Here, we vary the central density between the range $100$ and $3500$ {\rm MeV}/{\rm fm}$^3$.
The dotted section of each curve corresponds to the unstable configuration where $dM/d\epsilon_{c} < 0$.
The maxima of the mass-central density relations are
easily determined and then summarized in Table \ref{ta11} for
the EoSs investigated in this work.

Apart from the above discussion, here we need to further study the stability by examining adiabatic  index ($\gamma$) based on our EoSs concerning quark matter models. It is noted that   the adiabatic index is a basic ingredient of the  instability criterion, and is related to the thermodynamical quantity. Concerning the dynamical theory of infinitesimal, and adiabatic, radial oscillations of relativistic stars have been the first under investigation, more than 56 years ago by Chandrasekhar \citep{Chandrasekhar}.  The main conclusion regarding this study  is that the critical adiabatic index  $\gamma_{cr}$, for the onset of instability, increases due to relativistic effects from the Newtonian value $\gamma=4/3$.  For an adiabatic perturbation, the adiabatic index, which, for adiabatic oscillations, is related to the sound speed through and defined by 
 \citep{Chandrasekhar,Merafina}
\begin{equation}\label{adi}
\gamma \equiv \left(1+\frac{\epsilon}{P}\right)\left(\frac{dP}{d\epsilon}\right)_S,
\end{equation}
where  $dP/ d\epsilon$  is the speed of sound in units of speed of light and the subscript $S$ indicates at constant specific entropy.
 The Eq. (\ref{adi}) is the adiabatic index associated to the perturbations, or `effective'  \citep{Merafina}. Thus, the
`effective' must be greater than 
c for the configuration to be stable against radial
perturbations.

Thus, the `effective' $\gamma$ must be greater than  $\gamma_{cr}$ i.e. $\gamma > \gamma_{cr}$ 
for the configuration to be stable against radial perturbations. Following \citep{Moustakidis:2016ndw},
author has shown that this conditions are also applicable  to describe compact objects including white dwarf, 
neutron stars and supermassive stars. In \citep{Haensel},
the value of $\gamma$ lies between  2 to 4 for the EoS related to neutron star matter. Reference \citep{Glass} 
estimates the value of $\gamma> 4/3$ for relativistic polytropes depending on the ratio $\epsilon/P$ at the 
centre of the star. More fruitful discussion was found in \citep{Chavanis} for 
stability of an extended cluster with $\rho_e/ \rho_0 \ll 1$ in Newtonian gravity with $\gamma  > 4/3$.

The plots of $\gamma$ depending on the values of EoS parameters are shown in Fig. \ref{gamma_MLQ_CSS}.
It can be seen from Fig. \ref{gamma_MLQ_CSS} that our model 
is stable  against the radial adiabatic infinitesimal perturbations and increasing values of $\gamma$
mean the growth of pressure for a given increase in energy density.

%%%%%%%%%%%%%%%%%%%%%%%%%%%%%%%%%%%%%%%%%%%%
\section{Conclusions and astrophysical implications }\label{sec6}
%%%%%%%%%%%%%%%%%%%%%%%%%%%%%%%%%%%%%%%%%%%%
In this paper, we investigate the features of $4D$ Einstein-Gauss-Bonnet gravity in an extreme circumstances such as those arising within highly compact static spherically symmetric bodies. We considered the self-bound strange matter hypothesis. The interesting part of this theory is that the resulting regularized $4D$ EGB gravity has nontrivial dynamics and free from the Ostrogradsky instability.  

There exist considerable evidences that the possible existence of compact stars are partially or totally made up of quark matter. But the existence of quark stars is still controversial and its EoS is also uncertain. Here, we first considered the static spherically symmetric $D$-dimensional metric and derived corresponding field equations taking a limit of $D\rightarrow 4$ at the level of field equations. We then numerically solved field equations for strange matter hypothesis. To clarify the astrophysical implications of our work, we discuss two important scenarios. Firstly, we considered quark matter phases consisting of massless quarks, and secondly quark matter at zero temperature. 

To gain better understanding of the physical properties, we quantified the maximal mass from the central density and mass-radius relation of the stellar structure. The mass-radius results are graphically shown which strictly depends on the values of the coupling constant and the chosen EoS. Then, we showed that for $\alpha \to 0$  limit, the obtained TOV in $4D$ EGB gravity reduces to the standard Einstein theory, and solutions are compared in Table \ref{ta11}. Observing the Fig. \ref{f6}, we found that massless quark stars can achieve much higher masses and radii than cold star ones within the constraint of $\sim 2 M_\odot$. Since negative $\alpha$ reduces the maximum mass of a compact star for a given EoS. Furthermore, we obtain the interesting results of their physical properties such as the compactness and the corresponding effective adiabatic index, $\gamma$, which appears in the stability formula introduced by Chandrasekhar. We found that the obtained value of $\gamma> \gamma_{cr}$ for critical adiabatic index, for both equations of state considered here. In other words, the stability in all the cases  is ensured for quark matter EoSs.

Finally, it is notable that the investigation for other compact objects such as neutron star and white dwarf using the same context and its modified TOV equation are interesting subjects. However, we will leave these interesting topics for our future work.

\section*{Acknowledgments}
The authors are grateful to the referee for careful reading of the paper and valuable
suggestions and comments. TT would like to thank the financial support from the Science Achievement Scholarship of Thailand (SAST). P. Channuie acknowledged the Mid-Cereer Research Grant 2020 from National Research Council of Thailand under a contract No. NFS6400117.


\begin{thebibliography}{}

%%%%%%%%%%%%%% A %%%%%%%%%%%%%%%%%%%%%%%%%%%%
\bibitem[Alford et al.(2007)]{alford2007}
Alford M. G., Rajagopal K., Schaefer T. and Schmitt A.~2007, RMP, 80, 1455

\bibitem[Alford \& Rajagopal (2002)]{Alford2002}
Alford M. \& Rajagopal K.~2002, JHEP, 06, 031 

\bibitem[Alford (2004)]{Alford2004}
Alford M.~2004, PTPS, 153, 1

\bibitem[Alford et al.(2001)]{alf01}
Alford M., Rajagopal K., Reddy S. and Wilczek F.~2001, PRD, 64, 074017

\bibitem[Ali \& Mansoori (2020)]{sa03}
Ali, S. \& Mansoori, H.~2020, arXiv:2003.13382

\bibitem[Ai (2020)]{Ai:2020peo}
Ali, W.~Y.~2020, arXiv:2004.02858

\bibitem[Aragon et al.(2020)]{ar04}
Aragon A., Ramon B., Gonzalez P. A. and Vasquez Y.~2020, arXiv:2004.05632

\bibitem[Arbanil \& Malheiro (2016)]{Arbanil:2016wud}
Arbanil, J.~D.~V. \& Malheiro M.~2016, JCAP, 11, 012

\bibitem[Arrechea et al.(2020)]{Arrechea:2020evj}
Arrechea, J., Delhom A. and Jimenez-Cano A.~2020, arXiv:2004.12998

\bibitem[Aoki et al.(2020)]{Aoki:2020lig}
K.~Aoki, M.~A.~Gorji and S.~Mukohyama,
%``A consistent theory of $D \to 4$ Einstein-Gauss-Bonnet gravity,''
Phys. Lett. B~2020, 810, 135843
%%%%%%%%%%%%%%%%%%%%%%%%%%% END %%%%%%%%%%%%%%%%%%%%%%%%

%%%%%%%%%%%%%%%%%%%%%%%%%% B %%%%%%%%%%%%%%%%%%%%%%
\bibitem[Banerjee \& Singh (2020)]{Banerjee:2020stc}
Banerjee, A. \& Singh K. N.~2020, arXiv:2005.04028

\bibitem[Boulware \& Deser (1985)]{Boulware:1985wk}
Boulware, D. G. \& Deser, S.~1985, PRL, 55, 2656

%%%%%%%%%%%%%%%%%%%%%%%%% END B %%%%%%%%%%%%%%%%%%%%%%

%%%%%%%%%%%%%%%%%%%%%%%%%% C %%%%%%%%%%%%%%%%%%%%%

\bibitem[Callan et al.(1985)]{Callan:1985ia}
Callan C. G. Jr., Martinec E. J., Perry M. J. and Friedan D. ~1985, NPB, 262,  593

\bibitem[Churilova (2020)]{Churilova:2020aca}
Churilova, M. S.~2020, arXiv:2004.00513 

\bibitem[Cognola et al.(2013)]{Cognola:2013fva}
Cognola G., Myrzakulov R., Sebastiani L. and Zerbini S.~2013, PRD, 88, 024006

\bibitem[Casalino et al.(2013)]{Casalino:2020kbt}
Casalino A., Colleaux A., Rinaldi M. and Vicentini S.~2020, arXiv:2003.07068

\bibitem[Chandrasekhar (1964)]{Chandrasekhar}
Chandrasekhar, S.~1964, AJ, 140, 417

\bibitem[Chanmugan (1977)]{Chanmugan1977}
Chanmugan, G.~1977, ApJ 217, 799

\bibitem[Chavanis (2002)]{Chavanis}
Chavanis, P. H.~2002, Astron. Astrophys.  381, 709

\bibitem[Clifton et al.(2013)]{Clifton:2020xhc}
Clifton, T., Carrilho, P., Fernandes, P.~G.~S. \&  Mulryne, D.~J~2020, Phys. Rev. D, 102, 084005


%%%%%%%%%%%%%%%%%%%%%%%%% END C %%%%%%%%%%%%%%%%%%

%%%%%%%%%%%%%%%%%%%%%%% D %%%%%%%%%%%%%%%%%%%%%%

\bibitem[Doneva \& Yazadjiev (2020)]{Doneva:2020ped}
Doneva, D. D. \& Yazadjiev S. S.~2020, arXiv:2003.10284

%%%%%%%%%%%%%%%%%%%%%%%% END D %%%%%%%%%%%%%%%%%

%%%%%%%%%%%%%%%%%%%%%%%% E %%%%%%%%%%%%%%%%%%%%%

%%%%%%%%%%%%%%%%%%%%%%%% END E %%%%%%%%%%%%%%%%%

%%%%%%%%%%%%%%%%%%%%%%%% F %%%%%%%%%%%%%%%%%%%%%
\bibitem[Fernandes et al.(2020)]{Fernandes:2020nbq}
Fernandes, P.~G.~S., \textit{et al.}~2020, PRD, 102,  024025

\bibitem[Farhi \& Jaffe (1984)]{far84}
Farhi E. \& Jaffe R.L.~1984, PRD, 30, 2379

\bibitem[Fraga \& Romatschke (2005)]{Fraga:2004gz}
Fraga, E. S. \& Romatschke P.~2005, PRD, 71, 105014

%%%%%%%%%%%%%%%%%%%%%%%% END F %%%%%%%%%%%%%%%%%

%%%%%%%%%%%%%%%%%%%%%%%% G %%%%%%%%%%%%%%%%%%%%%

\bibitem[Ghosh \& Maharaj (2020)]{Ghosh:2020vpc}
Ghosh, S. G. \& Maharaj S. D.~2020, arXiv:2003.09841

\bibitem[Ghosh \& Kumar (2020)]{Ghosh:2020syx}
Ghosh, S. G. \& Kumar, R.~2020, arXiv:2003.12291

\bibitem[Glass \& Harpaz (1983)]{Glass}
Glass, E. N. \& Harpaz A.~1983, MNRAS, 202,  1

\bibitem[Glavan \& Lin (2020)]{Glavan:2019inb}
Glavan, D. \& Lin, C.~2020, PRL, 124, 081301

\bibitem[Glendenning (2000)]{Glendenning2000}
Glendenning, N. K. 2000, Compact stars (New York: Springer, Chapter 12.)

\bibitem[Glendenning (2000)]{Glendenning:2000dh}
Glendenning, N. K.~2000, PRL, 85, 1150

\bibitem[Glendenning \& Kettner (2000)]{Glendenning:1998ag}
Glendenning, N. K. \& Kettner C.~2000,  A \& A \textbf{353}, L9

\bibitem[Guo \& Li (2020)]{Guo:2020zmf}
Guo, M. \& Li, P. C.~2020, arXiv:2003.02523

\bibitem[Gurses et al.(2020)]{Gurses:2020ofy}
Gurses, M., Sisman T.~C. and Tekin B.~2020, arXiv:2004.03390

%%%%%%%%%%%%%%%%%%%%%%%% END G %%%%%%%%%%%%%%%%%

%%%%%%%%%%%%%%%%%%%%%%%% H %%%%%%%%%%%%%%%%%%%%%

\bibitem[Haensel et al.(2007)]{Haensel}
Haensel P., Potekhin A.Y. and Yakovlev D. G.~2007, Neutron Stars 1: Equation of State and Structure (Springer-Verlag, New York)

\bibitem[Haensel et al.(1986)]{Haensel:1986qb}
Haensel P., Zdunik J. and Schaeffer R.~1986, A\&A, 160, 121

\bibitem[Haensel et al.(2020)]{Haensel1986}
Haensel P., Zdunik J. L. and Schaefer R.~1986, A\&A, 160, 1 

\bibitem[Hansen(2020)]{Hansen:2013owa}
Hansen, D., \& Yunes, N.~2013
%``Applicability of the Newman-Janis Algorithm to Black Hole Solutions of Modified Gravity Theories,''
Phys. Rev. D, 88, 104020 

\bibitem[Hennigar et al.(2013)]{Hennigar:2020lsl}
Hennigar R. A., Kubiznak D., Mann R. B. and Pollack C.~2020, arXiv:2004.09472

\bibitem[Horndeski (1974)]{Horndeski:1974wa}
Horndeski, G.~W.~1974, Int. J. Theor. Phys. \textbf{10}, 363  

\bibitem[Heydari-Fard et al.(2020)]{Heydari-Fard:2020sib}
Heydari-Fard M., Heydari-Fard M. and Sepangi H. R.~2020, arXiv:2004.02140

%%%%%%%%%%%%%%%%%%%%%%%% END H %%%%%%%%%%%%%%%%%

%%%%%%%%%%%%%%%%%%%%%%%% I %%%%%%%%%%%%%%%%%%%%%

\bibitem[Islam et al.(2020)]{Islam:2020xmy}
Islam S. U., Kumar R. and Ghosh S. G.~2020, arXiv:2004.01038

%%%%%%%%%%%%%%%%%%%%%%%% END I %%%%%%%%%%%%%%%%%

%%%%%%%%%%%%%%%%%%%%%%%% J %%%%%%%%%%%%%%%%%%%%%

\bibitem[Jin et al.(2020)]{Jin:2020emq}
Jin X. H., Gao Y. X. and Liu D. J.~2020, arXiv:2004.02261

\bibitem[Jusufi (2020)]{Jusufi:2020qyw}
Jusufi, K.~2020, arXiv:2005.00360   

\bibitem[Jusufi et al.(2020)]{Jusufi:2020yus}
 Jusufi K., Banerjee A. and Ghosh S. G.~2020, arXiv:2004.10750 

%%%%%%%%%%%%%%%%%%%%%%%% END J %%%%%%%%%%%%%%%%%

%%%%%%%%%%%%%%%%%%%%%%%% K %%%%%%%%%%%%%%%%%%%%%

\bibitem[Kobayashi (2019)]{Kobayashi:2019hrl}
Kobayashi, T.~2019, Rept. Prog. Phys., 82,  086901 

\bibitem[Kobayashi (2020)]{Kobayashi:2020wqy}
Kobayashi, T.~2020, JCAP, 07, 013 

\bibitem[Konoplya \& Zhidenko (2020)]{Konoplya:2020juj}
Konoplya, R. A. \& Zhidenko, A.F.~2020, Phys. Dark Universe, 30, 100697 

\bibitem[Konoplya \& Zinhailo (2020)]{Konoplya:2020cbv}
Konoplya, R. A. \& Zinhailo, A. F.~2020, Phys.Lett.B, 810, 135793 


\bibitem[Kumar et al.(2020)]{Kumar:2020sag}
Kumar R., Islam S. U. and Ghosh S. G.~2020, arXiv:2004.12970

\bibitem[Kumar \& Ghosh (2020)]{Kumar:2020xvu}
Kumar, A. \& Ghosh, S. G.~2020, arXiv:2004.01131

\bibitem[Kumar \& Kumar (2020)]{Kumar:2020uyz}
Kumar, A. \& Kumar, R.~2020, arXiv:2003.13104

\bibitem[Kumar(2020)]{Kumar:2020owy} 
  Kumar, R., \& Ghosh, S.~G.~2020
  %``Rotating black holes in $4D$ Einstein-Gauss-Bonnet gravity and its shadow,''
  JCAP,  2007,  053 

%%%%%%%%%%%%%%%%%%%%%%%% END K %%%%%%%%%%%%%%%%%

%%%%%%%%%%%%%%%%%%%%%%%% L %%%%%%%%%%%%%%%%%%%%%

\bibitem[Lanczos (1938)]{Lanczos:1938sf}
Lanczos, O.~1938, Annals Math. 39, 842

\bibitem[Lin et al.(2020)]{Lin:2020kqe}
Lin, Z.~C. \textit{et al.}~2020, arXiv:2006.07913 [gr-qc]

\bibitem[Liu et al.(2020)]{Liu:2020vkh}
Liu C., Zhu T. and Wu Q.~2020, arXiv:2004.01662


\bibitem[Liu et al.(2020)]{Liu20201}
Liu P., Niu C. and Zhang C.-Y.~2020, arXiv:2005.01507

\bibitem[Liu et al.(2020)]{liu20}
 Liu P., Niu C., Wang X. and Zhang C.-Y.~2020, arXiv:2004.14267 

\bibitem[Lovelock (1971)]{Lovelock}
Lovelock, D.~1971, JMP, 12, 498

\bibitem[Lovelock (1972)]{Lovelock:1972vz}
Lovelock, D.~1972, JMP, 13, 874

\bibitem[Lu \& Pang (2020)]{Lu:2020iav}
Lu, H. \& Pang Y.~2020, Phys.Lett.B, 809, 135717

\bibitem[Lugones \& Horvath (2002)]{lugo02}
Lugones, G. \& Horvath J. E.~2002, PRD, 66, 074017

\bibitem[Liu et al.(2020)]{li04}
Liu, P., Niu, C. \& Zhang, C.-Y.~2020, arXiv:2004.10620 [gr-qc]


%%%%%%%%%%%%%%%%%%%%%%%% END L %%%%%%%%%%%%%%%%%

%%%%%%%%%%%%%%%%%%%%%%% M %%%%%%%%%%%%%%%%%%%%%%

\bibitem[Ma \& Lu (2020)]{Ma:2020ufk}
Ma, L. \& Lu H.~2020, arXiv:2004.14738 

\bibitem[Mahapatra (2020)]{Mahapatra:2020rds}
Mahapatra, S.~2020, EPJC, 80, 992 

\bibitem[Mishra (2020)]{Mishra:2020gce}
Mishra, A. K.~2020, Gen. Relat. Grav., 52, 106 

\bibitem[Merafina \& Ruffini (1989)]{Merafina}
Merafina, M., \& Ruffini, R.~1989, Astron. Astrophys. 221, 4

\bibitem[Moustakidis (2017)]{Moustakidis:2016ndw}
Moustakidis, C.~C.~2020, Gen. Rel. Grav. 49,  68

%%%%%%%%%%%%%%%%%%%%%%%% END N %%%%%%%%%%%%%%%%

\bibitem[Naveena(2020)]{NaveenaKumara:2020rmi} 
	Naveena, K. A., Rizwan, C.~L.~A.,  Hegde, K., Ali, M.~S., \& Ajit,~K.~M.
  %``Rotating 4D Gauss-Bonnet black hole as particle accelerator,''
  arXiv:2004.04521 [gr-qc].
%%%%%%%%%%%%%%%%%%%%%%%% R %%%%%%%%%%%%%%%%%%

\bibitem[Rajagopal \& Wilczek (2001)]{RAJAGOPAL2001}
Rajagopal, K. \& Wilczek F.~2001, arXiv:0011333

\bibitem[Rajagopal \& Wilczek (2001)]{raj01}
Rajagopal, K. \& Wilczek F.~2001, PRL, 86, 3492

%%%%%%%%%%%%%%%%%%%%% S %%%%%%%%%%%%%%%%%%


\bibitem[Samart \& Channuie (2020)]{Samart:2020sxj}
Samart, D. \& Channuie P.~2020, arXiv:2005.02826  

\bibitem[Steiner et al.(2002)]{Steiner:2002gx}
Steiner A. W., Reddy S. and Prakash M.~2002, PRD, 66, 094007

\bibitem[Schäfer \& Wilczek (1999)]{Schafer:1999jg}
Schäfer, T. \& Wilczek F.~1999, PRD, 60, 114033

\bibitem[shu (2020)]{shu}
Shu, F.~W.~2020, Phys. Lett. B, 811, 135907

%%%%%%%%%%%%%%%%%%%% T %%%%%%%%%%%%%%%%%

\bibitem[Tomozawa (2011)]{Tomozawa:2011gp}
Tomozawa, Y.~1986, arXiv:1107.1424

%%%%%%%%%%%%%%%%%%%% V %%%%%%%%%%%%%%%%%

\bibitem[Vath \& Chanmugam (1992)]{Vath1992}
Vath, H. M. \& Chanmugam G.~1992, A\&A, 260, 250-254

%%%%%%%%%%%%%%%%%%%%% W %%%%%%%%%%%%%%%%%%%

\bibitem[Wei \& Liu (2020)]{we03}
Wei, S.-W. \& Liu, Y.-X.~2020, arXiv:2003.07769

\bibitem[Wheeler (1986)]{Wheeler:1986}
Wheeler, J. T.~1986, NPB, 268, 737

\bibitem[Wiltshire (1985)]{Wiltshire:1985us}
Wiltshire, D. L.~1938, PLB, 169, 36

\bibitem[Witten (1984)]{Witten}
Witten, E.~1984, PRD, 30, 272

%%%%%%%%%%%%%%%%%%%% Y %%%%%%%%%%%%%%%%%%

\bibitem[Yang et al.(2020)]{Yang:2020jno}
 Yang K., Gu B. M., Wei S. W. and Liu Y. X.~2020, arXiv:2004.14468 
 
\bibitem[Yang et al.(2020)]{si20}
 Yang S.-J., Wan J.-J., Chen J., Yang J. and Wang Y.-Q.~2020, arXiv:2004.07934 

%%%%%%%%%%%%%%%%%%% Z %%%%%%%%%%%%%%%%%

 \bibitem[Zeng et al.(2020)]{Zeng:2020dco}
Zeng X. X., Zhang H. Q. and Zhang H.~2020, arXiv:2004.12074

\bibitem[Zhang et al.(2020)]{Zhang:2020qam}
 Zhang C. Y., Li P. C. and Guo M.~2020, arXiv:2003.13068

\bibitem[Zhang et al.(2020)]{Zhang:2020sjh}
Zhang C. Y., Zhang S. J., Li P. C. and Guo M.~2020, arXiv:2004.03141

\bibitem[Zhang et al.(2020)]{zh03}
 Zhang Y.-P., Wei S.-W. and Liu Y.-X.~2020, arxiv:2003.10960

\bibitem[Zwiebach (1985)]{Zwiebach:1985uq}
Zwiebach, B.~1985, Phys. Lett.,  B156, 315

\end{thebibliography}
\end{document}